\documentclass[aps,prb,reprint,showpacs,showkeys,superscriptaddress,preprintnumbers,amsmath,amssymb]{revtex4-1}
\usepackage{multirow}
\usepackage{graphicx}
\usepackage{bm}
\usepackage{dcolumn}
\usepackage{picture}
\usepackage[hidelinks]{hyperref}
\usepackage{natbib}
\usepackage{amsmath}
\usepackage{times}
\usepackage{color}
\usepackage{xcolor}
\usepackage{soul}
\usepackage{amsmath}
\usepackage{amssymb}

\begin{document}
	\title{Dilute stuffing in the pyrochlore iridate Eu$ _2 $Ir$ _2 $O$ _7 $}
	\author{Prachi Telang}
	\affiliation{Department of Physics, Indian Institute of Science Education and Research, Pune, Maharashtra-411008, India}
	
	\author{Kshiti Mishra}
	\affiliation{Department of Physics, Indian Institute of Science Education and Research, Pune, Maharashtra-411008, India}
	
	\author{A. K. Sood}
	\affiliation{Department of Physics, Indian Institute of Science, Bangalore-560012, India}
	
	\author{Surjeet Singh}
	\email[email:]{surjeet.singh@iiserpune.ac.in}
	\affiliation{Department of Physics, Indian Institute of Science Education and Research, Pune, Maharashtra-411008, India}\affiliation{Center for Energy Science, Indian Institute of Science Education and Research, Pune, Maharashtra-411008, India}

\date{\today}

\begin{abstract}
The pyrochlore Eu$_2$Ir$_2$O$_7$ has recently attracted significant attention as a candidate Weyl semimetal. The previous reports on this compound unanimously show a thermally induced metal to insulator (MI) transition, concomitant with antiferromagnetic (AFM) long-range ordering of the Ir-moments below T$_\textit{N} \sim  $120 K. However, there are contradictory reports concerning the slope d$\rho/$dT of the resistivity plots ($\rho$) in the "metallic" state above the metal-insulator (MI) transition, and the value of $\rho$ in the insulating state, both of which show significant sample dependence. Here, we explore this issue by investigating six different Eu$_2$Ir$_2$O$_7$ samples with slightly varying Eu:Ir ratio. High-resolution synchrotron powder diffraction are done to probe minor variations in the cell parameters of the various Eu$_2$Ir$_2$O$_7$ samples investigated here. Specific heat (C$ _p $) and magnetic susceptibility of all the samples showed long-range antiferromagnetic ordering upon cooling below T$ _\textit{N} \sim $120 K. The transitions are, however, found to be smeared out for the off-stoichiometric samples. We show that the sign of d$\rho/$dT above the metal-insulator (MI) transition is highly sensitive to the unit cell length, which, in turn, depends on the level of Eu-stuffing at the Ir-site. Samples with composition close to the ideal stoichiometry (Eu : Ir $ = $ 1) showed a change of sign of  d$\rho/$dT from negative to positive upon cooling below a certain temperature T $^*$ $>$ T$_\textit{MI}$. With increasing Eu-stuffing T$ ^* $ decreased until a negative d$\rho/$dT persisted without any sign change down to T$_\textit{MI}$.     

\end{abstract}

\pacs{}

\maketitle

\section{Introduction}
\label{Intro}
The geometrically frustrated pyrochlore (A$_2$B$_2$O$_7$) structure is well-known to host several exotic quantum many-body ground states, including quantum spin liquids and spin ices \cite{Gardner, Balents}.  In some pyrcohlore oxides with a 5\textit{d} transition metal ion (e.g., Ir$ ^{4+} $) at the \textit{B}-site, presence of a relativistic spin-orbit (SO) interaction term in the Hamiltonian gives rise to novel topological phases not present in their \textit{3d} and \textit{4d} analogues. For this reason, the iridates of the pyrochlore structure have gained significant attention in the recent years \cite{WanWeyl, Krempa, Hongbin, Millis}. In the absence of SO interaction, an Ir$ ^{4+} $ ion in the octahedral coordination of the pyrochlore structure will be  in a 5d$^5$ ($t^5_{2g} $ $e^0_g$) ground state. However, due to SO interaction, which is sizable for an Ir$ ^{4+} $ ion,  the $t_{2g} $ level splits further into a completely-filled quadruplet (J$ _\textit{eff} = 3/2 $), and a higher-lying half-filled doublet (J$ _\textit{eff} = 1/2 $), giving rise to an effective J$ _\textit{eff} = 1/2 $ moment on the frustrated pyrochlore lattice.  

The physical behavior of pyrochlores iridates changes from that of an antiferromagnetic-insulator for smaller or heavier  rare-earths (i.e., A = Gd, Tb, Dy, Ho, Er and Yb, including Y) to an exotic non-magnetic metal for Pr$_2$Ir$_2$O$_7$ (Ref. \citenum{Tokiwa}). The intermediate members corresponding to R = Nd, Sm and Eu show a thermally induced metal-insulator transition (MI) upon cooling below a temperature (T$ _\textit{MI}$), concomitant with the onset of antiferromagnetic (AFM) long-range ordering  of Ir$ ^{4+} $ moments at T$_N$ (Ref. \citenum{Matsuhira2011}). These intermediate members are proposed to stabilize a novel topological phase called the Weyl semimetal \cite{WanWeyl}. Among these, the member Eu$ _2 $Ir$ _2 $O$_7$ (hereafter, abbreviated as EIO), has attracted particular attention because not only it shows a robust MI transition, it also offers the advantage of a non-magnetic A-site since in Eu$ ^{3+} $ ion the spin (S) and orbital angular momenta (L) mutually compensate each other. Therefore, the magnetism and associated topological properties arises \textit{solely} due to the 5d electrons of Iridium. An evidence of the Weyl semimetallic state in EIO has been recently claimed in ref. \citenum{shushkov} using the THz optical conductivity studies.

Upon cooling below T$ _\textit{MI}$, the resistivity ($ \rho$) of EIO increases sharply, the state below this temperature is therefore dubbed as the 'insulating' state. On the other hand, in the 'metallic' state above T$ _\textit{MI}$, $\rho(T)$ exhibits an anomalous behavior with the slope d$\rho$(T)/dT negative in some reports\cite{Zhao, Takatsu, Matsuhira2007, Matsuhira2011} and positive in others\cite{Ishikawa, Clancy}. However, what causes to this stark sample dependence of d$\rho$(T)/dT in the 'metallic' state has not been properly investigated to the best of our knowledge. A high-pressure investigation by Tafti et al. \cite{Tafti} shows that the sign of d$\rho(T)$/dT above T$ _\textit{MI}$ depends sensitively on the external pressure changing its sign from negative under ambient or low pressures to positive above 6 GPa.  

Due to the ionic-size mismatch between Eu$ ^{3+} $ and Ir$ ^{4+} $, a minor Eu-Ir off-stoichiometry in a nominally stoichiometric EIO sample may also induce a small chemical pressure, the question is will this pressure be large enough to explain the sign of d$\rho$(T)/dT in various EIO samples previously reported? However, a systematic study of cation anti-site disorder, which is also commonly referred to as 'stuffing' (for A-site ion occupying  the B-site) or 'negative stuffing' (when the reverse happens) in the pyrochlore literature, is lacking for the iridiate pyrochlores. In the recent year, such studies on the insulating titanate pyrochlores unveiled several interesting aspects of their ground state properties\cite{Gd2Ti2O7,Baroudi,Rossstuffedtitanates}. For example dilute Yb stuffing in the pyrochlore Yb$_2$Ti$_2$O$_7$ is reported to have a significant effect on the magnetic ground state where the ferromagnetic T$_c$ is suppressed by nearly 25\%  by a mere 2\% of stuffing, and the shape of the specific heat anomaly also changes considerably. Therefore, it is desirable to investigate and understand the effect of stuffing in the iridate pyrochlores; particularly in the candidate Weyl semimetal EIO where the slope $\rho$(T) versus T curve exhibits a significant sample dependence.  

Here, we investigate the effect of minor off-stoichiometry on the physical properties of EIO. Specific heat (C$_p$ ), resistivity ($ \rho $), thermoelectric power (S) and magnetic susceptibility ($ \chi $) are studied for \textit{six} different EIO samples with slightly varying stoichiometries. Changes in the physical properties due to off-stoichiometry are correlated with  minor changes in the lattice parameter measured using a very high-resolution synchrotron powder X-ray diffraction data. We show that EIO samples synthesized in air forms with varying levels of Eu-stuffing at the Ir-site. The extent of Eu stuffing depends not as much on small variations in the starting Eu : Ir stoichiometry as on the synthesis parameters, including the highest temperature employed and the duration of synthesis at this temperature. We show that stuffing results in a very small unit cell expansion, giving rise to a negative chemical pressure enough to change the sign of slope of $\rho(T)$ curve above T$_\textit{MI}$. We establish a clear correlation between the unit cell volume change due to stuffing and  the sign of d$\rho$/dT. Stuffing also affects the magnitude of $\rho$ due to charge carrier doping, and tends to broaden and separate out the AFM/MI transition. 

The rest of the manuscript has been organized as follows: Experimental details including sample synthesis are given in section \ref{ExpDet}, which is followed by results \& discussion in section \ref{Result} under which we present the structure refinement in section \ref{xture} and physical properties in section \ref{properties}. A discussion on the sign of d$\rho$/dT appears under section \ref{disc}, which is followed by the concluding remarks and summary in section \ref{conc}.       

\begin{figure}[t]
	\centering
	\includegraphics[width=0.45\textwidth]{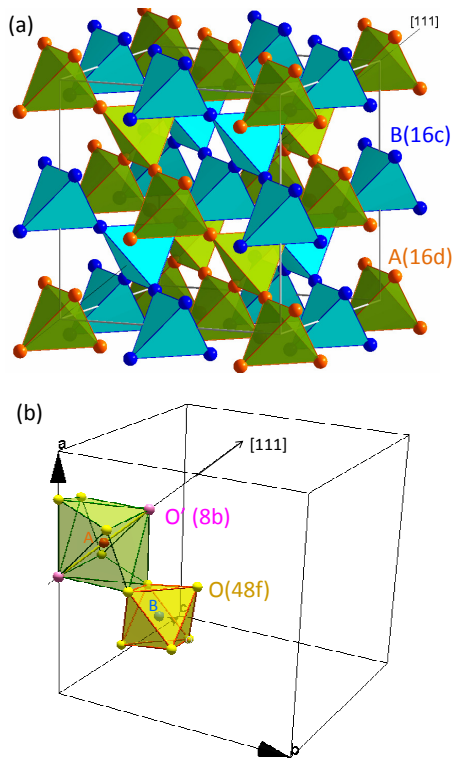}
	\caption {Conventional unit cell of the pyrochlore A$ _2 $B$ _2 $O$ _7 $ structure  with (a) only A$ ^{3+} $ and B$^{4+} $ sublattices shown, and (b) B-O$_6 $ octahedron and A-O$_6$-O$'_2$ scalenhedron are shown in the unit cell.}
	\label{structure}
\end{figure} 

\section{Experimental Details} 
\label{ExpDet}
Six different EIO samples, labeled as A, B1, B2, C1, C2 and D were prepared using the precursors  Eu$_2$O$_3 $ (Sigma Aldrich, 99.9 \%) and  IrO$_2 $ (Sigma Aldrich, 99.9 \%). The ratio Eu/Ir in the starting composition was varied as: Eu/Ir $ =  (1 - s)/(1 + s)$ where $ s  = 0$ (A), $ s = 0.01$ (B1), $ s = - 0.01$ (B2), $ s = 0.02$ (C1), $ s = - 0.02$ (C2) and $s = 0.015$ (D). As opposed to samples A, B1/B2 and C1/C2 that were synthesized in air, sample D was partly treated under vacuum as described later. The reactants were weighed with a precision of 0.1 mg. After weighing they were thoroughly ground together in an agate mortar and pestle. Subsequently, the mixtures were cold-pressed in a 13 mm stainless steel die under a pressure of 1500 Kg cm$^{-2} $. The pellets were fired in air at temperatures ranging from 800 up to 1070 $ ^\circ$C in the following sequence. 

Batch 1: Sample A was synthesized first by sintering for a total of 374 hrs with almost 20 intermediate grinding/cold-pressing cycles, and by progressively increasing the temperature in subsequent cycles. To prevent loss of volatile IrO$ _2 $, the temperature increment between any two successive firing cycles was never allowed to exceed 10 $ ^\circ$C; and for the same reason only 10 \% of total synthesis time was used for sintering at temperatures higher than 1030 $ ^\circ$C.    

Batch II: In batch II samples B1 and B2 were synthesized using the same protocol as used for A but in this case the highest sintering temperature employed was reduced to 1000 $ ^\circ$C.   

Batch III:: Samples C1 and C2 were synthesized in batch III. In this case, due to higher level of off-stoichiometry, the precursor materials took longer to react fully. Altogether, these compositions were sintered for 574 hrs with over 43 intermediate grinding/cold-pressing cycles at temperatures less than 1070 $ ^\circ$C. 

Batch IV: Sample labeled D was synthesized separately in bach IV with 20 intermediate grinding/cold-pressing cycles for a total of 320 hrs at temperatures less than 1030 $ ^\circ$C in air. At this point, the unreacted Eu$_2$O$_3$ and IrO$_2$ peaks were still present. Instead of sintering it further in air at higher temperatures, as done in the case of batch III samples, here an alternate route was employed wherein the sample was annealed at 1100 $ ^\circ$C for 60 hrs under high vacuum. This was done to suppress loss of Ir during sintering to induce negative stuffing as discussed later. The product obtained at the end of this treatment was reground and palletized, and subjected to a final sintering at 1000 $^\circ$C for 12 hrs. 

The synthesis process in each case was called to an end only after the powder X-ray diffraction indicated the formation of the pyrochlore phase with minor or no traces of diffraction peaks due to the precursor materials. The pyrochlore phase forms faster at higher temperatures but at the cost of volatile IrO$_2 $  loss from the sample that tends to sublimate excessively at temperatures exceeding $\approx$1050 $^\circ$C owing to its high vapor pressure. Normally, this loss can be quite significant if the sample is not sintered for a long enough duration at lower temperatures to react IrO$_2 $ with Eu$_2 $O$_3$.  

The phase formation during the sintering process was monitored using a Bruker D8 Advance powder X-ray diffractometer. The structural parameters were quantitatively refined using a very high resolution data collected at the powder diffraction beamline (MCX) of the ELETTRA synchrotron radiation facility, Trieste, Italy. For this purpose a Huber 4-axis X-ray diffractometer equipped with a fast scintillator detector was used\cite{Rebuffi}. The sample was prepared  in the form of finely grounded powder that was placed in glass capillary tube of inner diameter 0.1 mm. During the experiment the capillary was rotated at an angular speed of 180 rpm. Diffractograms were collected in the range 10$^\circ \leq2\theta \leq$ 46$^\circ$ with a step$\-\ $size of either 0.005$^\circ$ (A and D) or  0.01$^\circ$ (B1, B2, C1 and C2), and a counting time of 1s at each step. The incident beam energy was set at $19.7$ keV ($\lambda=0.6294 $\r{A}). The structure refinement was done by the Rietveld method using the \textit{FullProf} software\cite{RODRIGUEZCARVAJAL199355}. 

The chemical composition of the prepared samples was analyzed using the energy dispersive X-ray analysis (EDX) technique in a Zeiss Ultra Plus scanning electron microscope. Temperature and field-dependent magnetization measurements were done in the temperature range 2-300 K using a Magnetic Property Measurement System (MPMS) from Quantum Design (USA). Measurements were carried out under an applied magnetic field of 1 kOe in, both, field cooled (FC), and zero field cooled (ZFC) conditions. The resistivity and thermoelectric power was measured using a Physical Property Measurement System (PPMS) from Quantum Design (USA). The current and voltage contacts on the sample surface were made using a conducting silver epoxy and gold wire. Specific heat measurements were performed using the relaxation method in the PPMS. The heat capacity of the sample holder and APIEZON N grease (addenda) was determined prior to the measurements. High-resolution thermogravimety (TGA) was done using Mettler Toledo (TGA/DSC 3+) with sub-microgram weight resolution.

\section{Results \& discussion}
\label{Result}

\begin{figure*}[!]
	\centering
	\includegraphics[width=0.9\textwidth]{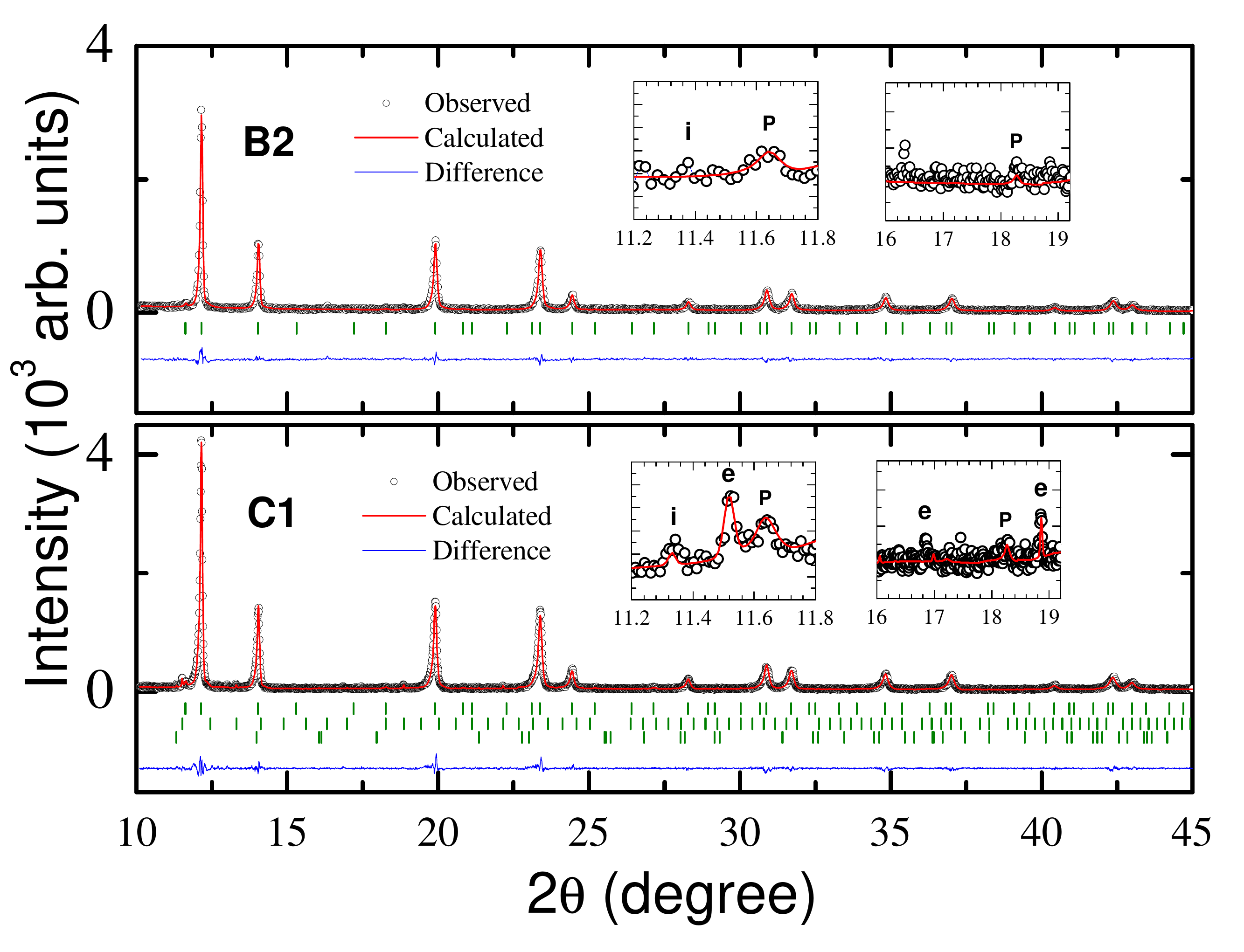}
	\caption {Rietveld refinement of synchrotron powder X-ray diffraction data of Eu$ _2 $Ir$ _2 $O$ _7 $ samples B2 and C1 (see text for details).Insets show the absence/presence of the secondary phases labeled as e $\equiv $ Eu$_2$O$_3$, i $\equiv $ Ir$O_2$ and * $\equiv $ unknown phase in B2 and C1, respectively. The vertical green bars below the diffraction pattern indicate the positions of the Bragg peaks of the pyrochlore phase Eu$ _2 $Ir$ _2 $O$ _7 $. In the lower panel where a multi-phase refinement was carried out, the additional second and third rows of vertical bars represent the positions of Bragg peaks due to Eu$ _2 $O$ _3 $ and IrO$ _2 $.}
	\label{xrd}
\end{figure*}

\subsection{Structural Characterizations}
\label{xture}
\begin{table*}
	\caption{Structural parameters for various Eu$_2$Ir$_2$O$_7$ samples, labeled A, B1, B2, C1, C2 \& D, obtained using Rietveld refinement of the high-resolution synchrotron powder X-ray diffraction data. Parameters characterizing the quality of fit are also included}
	\label{Table1}
	\begin{center}
		\begin{tabular}{c  c  c  c  c  c  c }
			\hline
			\hline
			\vspace*{-2mm}
			& & & &\\	
			Eu$_2$Ir$_2$O$_7$ samples  \hspace*{1cm} & $A$  \hspace*{0.3cm} & $B1$ \hspace*{0.3cm} & $B2$ \hspace*{0.3cm} & $C1$ \hspace*{0.3cm}& $C2$ \hspace*{0.3cm}& $D$  \\
			\vspace*{-2mm}
			& & & &\\	
			\hline
			\hline
			\vspace*{-3mm}
			& & &  &\\
			\vspace{1mm}
			\hspace*{-1.9cm}$a$(\AA)\hspace*{-0.4cm}& $10.2989$\hspace*{0.6cm} & $10.2994$\hspace*{0.6cm} & $10.3034$\hspace*{0.6cm}  & $10.3058$\hspace*{0.6cm} & $10.3039$\hspace*{0.6cm}  & $10.2965$ \\ 
			\vspace{1mm}
			\hspace*{-1.7cm} $u_{O(\textit{48f})}$\hspace*{-0.2cm}& $0.3370(3)$\hspace*{0.6cm} & $0.3432(4)$\hspace*{0.6cm} & $0.3399(4)$\hspace*{0.6cm} & $0.3345(4)$ \hspace*{0.6cm} & $0.3339(4)$\hspace*{0.6cm} & $0.3371(7)$ \\
			\vspace{1mm}
			\vspace{1mm}
			\hspace*{-1.7cm}Eu : Ir\hspace*{-0.2cm}& 1.001(0.013)\hspace*{0.6cm} & 1.002(0.014)\hspace*{0.6cm} & 1.011(0.010)\hspace*{0.6cm} & 1.049(0.009) \hspace*{0.6cm} & 1.002(0.010) \hspace*{0.6cm} & 1.009(0.026)\\  
			\vspace{1mm}
			\hspace*{-1.7cm}\textit{No.of phases refined}\hspace*{-0.2cm}& 2\hspace*{0.6cm} & 1\hspace*{0.6cm} & 1\hspace*{0.6cm} & 3\hspace*{0.6cm} & 2 \hspace*{0.6cm} & 3 \\
			\vspace{1mm}
			\hspace*{-1.7cm}\textit{Secondary Phase(s)} \hspace*{-0.2cm}& $Eu_2O_3$\hspace*{0.6cm} & Nil\hspace*{0.6cm} & Unidentified phase \hspace*{0.6cm} & Eu$ _2 $O$_3$, IrO$_2$ \hspace*{0.6cm} & Eu$ _2 $O$_3$ \hspace*{0.6cm}& Eu$ _2 $O$_3$, Ir-metal \\   
			\vspace{1mm}
			\hspace*{-1.7cm}$\chi^2$\hspace*{-0.2cm}& $2.06$\hspace*{0.6cm} & $1.14$\hspace*{0.6cm} &  $1.23$\hspace*{0.6cm} & $1.76$ \hspace*{0.6cm} & $1.51$ \hspace*{0.6cm}& $1.16$  \\
			\hspace*{-0.9cm} $R_p$ \hspace*{0.6cm} & $8.47$\hspace*{0.6cm}& $9.03$\hspace*{0.6cm} &  $8.87$\hspace*{0.6cm} & $9.96$\hspace*{0.6cm} & $10.02$\hspace*{0.6cm}   & $12.54$ \\
			\vspace{1mm}
			\hspace*{-1cm}$R_{wp}$\hspace*{0.6cm}& $11.2$\hspace*{0.6cm} &  $12$\hspace*{0.6cm} &   $12.1$\hspace*{0.6cm} &  $13.1$\hspace*{0.6cm} &  $13.3$\hspace*{0.6cm}  &  $13.4$ \\
			\vspace{1mm}
			\hspace*{-1cm}$R_e$\hspace*{0.6cm}& $7.77$\hspace*{0.6cm} & $11.1$\hspace*{0.6cm} &  $10.1$\hspace*{0.6cm} & $10.1$\hspace*{0.6cm} & $10.2$\hspace*{0.6cm}  & $9.8$\\ 
	\end{tabular}
	\end{center}
\end{table*}

Eu$_2$Ir$_2$O$_7$ or more suitably Eu$_2$Ir$_2$O$_6$O', which reflects the two inequivalent O-sites, crystallizes with the pyrochlore structure (space group: \textit{Fd-3m}, z = 8) with Eu, Ir, O and O' located, respectively, at \textit{16d} (0.5, 0.5, 0.5), \textit{16c} (0, 0, 0), \textit{48f} (\textit{u}, 0.125, 0.125) and \textit{8b} (0.375, 0.375, 0.375), positions. The cations Eu and Ir form interpenetrating corner-linked tetrahedral networks running parallel to the 111 direction of the cubic unit cell, as shown in Fig. \ref{structure}a. The two sublattices are displaced relative to one another along the unit cell edge by a length \textit{a}/2, where \textit{a} is the lattice parameter. In the pyrochlore structure, each B-site cation is coordinated to six O (\textit{48f}) ions, forming a BO$_6$ octahedron (Fig. \ref{structure}b). Since the \textit{48f} position has a variable x-coordinate, the symmetry of BO$_6$ octahedron depends on the value of \textit{u}. The perfect octahedral symmetry is achieved for \textit{u} = 0.3125. In interval $0.3125  \le u \le  0.375$, which defines the field-of-stability of the pyrochlore structure, the IrO$_6$ octahedral exhibits a trigonal compression that increase with \textit{u}. In most known pyrochlores, the value of \textit{u} lies well within these limits. In the pyrochlore iridates, in particular, the typical value of \textit{u} is reported to be around 0.333, and the corresponding value of the bond angle Ir$ - $O$ - $Ir varies from $\approx$127$^\circ $ to 130$^\circ $ (Ref. \citenum{SUBRAMANIAN198355})

The A-site ion in the pyrochlore structure forms an axially compressed scalenohedron with six equidistant $ 48f $ oxygens forming a puckered hexagonal A-O$ _6 $ ring; and two axially located O(\textit{8b}) that forms a $ 180^\circ $ O'$ - $A$ - $O' bond oriented perpendicular to the average plane of the A-O$ _6 $ ring. The scalenohedron becomes a perfect cube for \textit{u}  = 0.375. For smaller values of \textit{u} than 0.375, the bond length A$ - $O' becomes smaller than A$ - $O, compressing the scalenohedron axially. A sense of the relative orientations of the A- and B-site coordination polyhedrons can be gathered from the fact that bond O'$ - $A$ - $O' always points along the principal diagonal as shown in Fig. \ref{structure} irrespective of the values \textit{u} and \textit{a} take.

Because of high neutron absorption cross-section of Eu and Ir, structural determination using neutron scattering data, which would have given a more accurate information of the O(48f) positional parameter, is not possible in EIO.Hence minor variations in the unit cell parameters of our variously treated EIO samples were studied using the high resolution synchrotron powder X-ray diffraction technique. The disadvantage of X-ray is that it is not very sensitive to the position of lighter oxygen ion. As a result, the parameter \textit{u}, which decides the position of O(\textit{48f}) ion, remains vaguely defined. However, this problem is partly overcome using the high resolution synchrotron radiation and a point detector that allows for determination of lattice parameter to a very high precision. 

The high-resolution MCX data for each sample are analyzed using the Rietveld refinement. In Fig. \ref{xrd}, match between experimental and calculated pattern is shown for two representative samples. The observed diffraction patterns in each case can be well-fitted to the pyrochlore structure. Very weak diffraction peaks due to unreacted precursors and/or due to Ir-metal were also observed. In such cases, a mixed-phase refinement was carried out. The total amount of impurity phases in our samples varied between 1 and 2 \%. The data were fitted using the Thompson-cox-Hastings pseudo-Voigt line profile function to account for the slight peak asymmetry. Absorption correction was also taken into account, which is important for the samples with heavier elements \cite{Rebuffi}. The refinement was considered to have converged when the shifts in the parameters being refined became less than 10\% of their estimated standard deviation. The lattice constant (\textit{a}), variable O-position parameter (\textit{u}), isotropic thermal parameters (B) and occupancies of Eu and Ir were treated as variables. Eu and Ir occupancies were refined by imposing the constraint that sum of occupancies at $ 16c $ and $16d $ sites should be equal to 2. Such a constraint is commonly used in structural studies on stuffed pyrochlores \cite{Clancy}. Since oxygen is not very sensitive to X-ray scattering, the occupancies of both the O-sites were fixed as fully occupied. The main results of the Rietveld refinement are collected in Table.\ref{Table1}. A good quality of the refinement can be inferred from the fitted and the difference plots in Fig. \ref{xrd}. A moderately low values of the goodness-of-fit ($\chi^2$), and of the R-factors ($R_wp$, $R_p$ and $R_e$) reflects a satisfactory fit in each case.

\begin{figure}[!]
	\centering
	\includegraphics[width=0.45\textwidth]{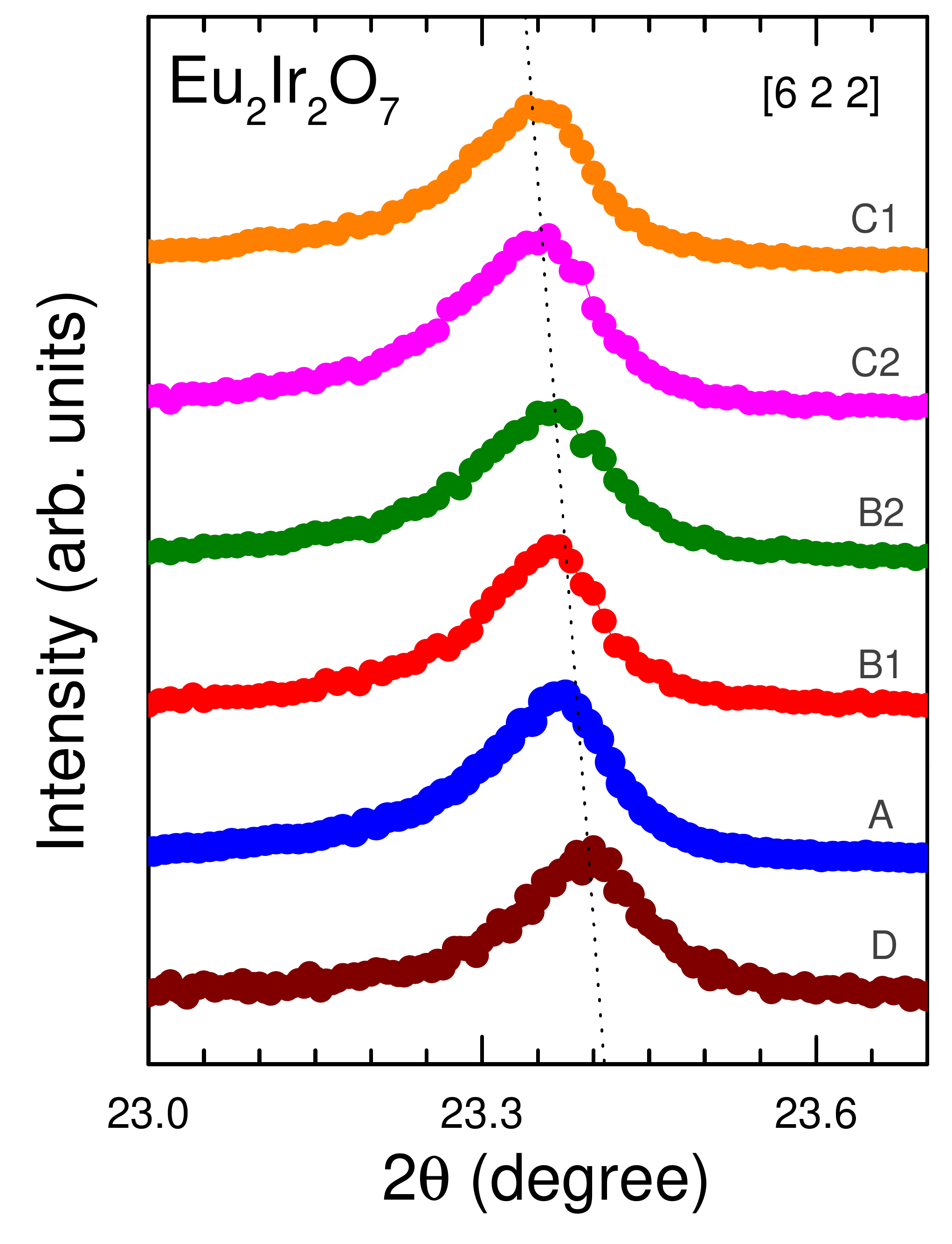}
	\caption{The high-resolution synchrotron powder X-ray diffraction data of Eu$_2$Ir$_2$O$_7$ samples A, B1, B2, C2, C1 and D (see table. \ref{Table1} for details) in the 2$\theta$ range covering the Bragg peak (6 2 2), chosen as representative to demonstrate a slight rightward shift of the pattern from C1 to A0. dashed line is a guide to eye.}
	\label{peakshift}
\end{figure} 

As shown in table I, the lattice parameter of our samples, obtained for each sample from the Rietveld refinement of the x-ray diffraction data for $2\theta$ = 10$^\circ$ to  46$^\circ$ , vary slightly, which is related to the Eu-stuffing discussed further. The average value of the lattice parameter of our samples is 10.302 \AA, which agrees fairly nicely with the same value reported by Chien and Sleight for their air synthesized sample\cite{Chien}. This value is, however, bigger than \textit{a} = 10.274 \AA, for single crystal specimens grown using KF flux \cite{Millican}. This difference can be attributed to negative stuffing (i.e., Ir occupying the Eu site), as argued by Ishikawa et al. \cite{Ishikawa}. In the present study, only sample D, that was treated partly under vacuum  is probably negatively stuffed and accordingly it has the smallest lattice parameter (\textit{a} = 10.297 \AA). The \textit{u}-parameter in our samples lies in the interval between 0.33 and 0.34. The error bar on the value of \textit{u}, listed in table \ref{Table1}, is taken directly from the output file of our Rietveld refinement; the actual standard deviation is expected to be much larger due to insensitivity of X-ray to the oxygen position. We shall, therefore, content ourselves with the average value of \textit{u} which is, $ \sim $ 0.334, in a fairly good agreement with similar values previously reported \cite{SUBRAMANIAN198355}. The actual variations in the value of \textit{u} of our samples due to stuffing is, at any rate, not expected to be very significant since the overall change in the lattice parameter itself is rather small. The average values of Ir$ - $O(\textit{48f}) bond length and Ir$ - $O(\textit{48f})-Ir bond angle for our various EIO samples is around 2.03 \AA, and nearly $\sim$127 $^\circ $, respectively. Both these values are comparable to the values previously reported for other iridate pyrochlores \cite{KENNEDY1997303}. 

A comprehensive neutron crystallographic refinement of the stuffed Yb2Ti2O7 samples was carried out by Ross et al. [18]. They established that the remarkable sample dependence of the magnetic ground state of this pyrochlore with the level of Yb stuffing at the Ti-site. In this work, the authors proposed that the room temperature lattice parameter could, in principle, be used to estimate the level of stuffing. Using this as a guideline, we associate the observed variation in the lattice parameter in our various EIO samples with the level of Eu-stuffing at the Ir-site, which appears to be valid given that the Eu : Ir ratio in our samples scales with the lattice parameter as shown in Table I.  We now examine the observed variation in the lattice parameters of our samples more closely. In Fig. \ref{peakshift}, the shift in position (2$\theta$) of the Bragg peak (6 2 2) obtained using the synchrotron data is shown for all the EIO samples. From here it is evident that the lattice parameter of sample C1 is the largest and that of D the smallest. We believe that this variation is due to the stuffing, analogous to reports of change in lattice parameter due to minor stuffing in the pyrochlore titanates \cite{Lau, Baroudi}. Since the ionic radius of Eu$^{3+}$ (1.066 \AA) is bigger than that of $Ir^{4+}$ (0.62  \AA) (Ref. \citenum{Shannon:a12967}), stuffing of Eu$^{3+}$ at the Ir-site results in an expansion of the lattice. On the other hand, a small level of negative stuffing causes the lattice to contract slightly. In the refined structural data, extent of Eu-stuffing is shown in table \ref{Table1}. Samples C1 and C2 are Eu-stuffed beyond the measurement errors. On the other hand, in samples A, B1, B2 and D, the level of stuffing, if any, is less than $\pm$ 1 \% which could not be any better resolved. However, from the trends in variation of lattice parameter, and judging from the physical properties presented latter, sample A and B1 appear to be closest to the ideal stoichiometry; and sample D appears to be slightly negatively stuffed. The most striking thing to note is that the initial or the starting composition (i.e., Eu/Ir in the starting mixture) is not as important in deciding the final stoichiometry as the details of the synthesis protocol. This statement is further testified by the physical properties that are nearly the same for samples prepared in a given batch, despite differences in their starting compositions.     

The extra phase(s), if any, in all the samples are recorded in the table \ref{Table1}. In sample A, where the starting composition was stoichiometric, some loss of IrO$ _2 $ during the synthesis may have resulted in traces of unreacted Eu$ _2 $O$ _3 $ in the final product. In sample B1, on the other hand, 1 \% excess IrO$ _2 $ in the starting mixture probably compensated for this loss, resulting in phase pure diffraction pattern with ratio Eu/Ir closer to the ideal value. Sample D was sintered under vacuum after a preliminary reaction of the precursors in air at T $ \le$ 1030 $ ^\circ C $. Upon sintering under vacuum for about 60 hrs at 1100 $ ^\circ C $, the final product had the desired pyrochlore phase as the main product but along with that minor diffraction peaks revealing the presence of Eu$_2$O$_3$ and Ir metal were also detected. This is the only sample where Ir-metal peaks were present, which is perhaps the \textit{main} drawback of sintering under vacuum. The advantage, however, is that by sintering under vacuum iridium losses can be minimized. It has been reported that at high temperatures, IrO$_2$ absorbs oxygen from its surrounding atmosphere to form a highly volatile oxide of iridium, namely, IrO$_3$ (Ref. \citenum{Cordfunke}). Thus sintering under inert atmosphere or vacuum prevents Ir losses by suppressing the IrO$_3$ formation. This is apparently the reason why vacuum sintered samples or those obtained from KF flux tend to show slight Ir excess in their measured stoichiometry \cite{Millican, Ishikawa}.  

\subsection{Physical Properties}
\label{properties}
We will next examine the specific heat (C$_p$), resistivity ($ \rho $), thermopower (S) and magnetization (M) of the various EIO samples to examine how minor changes in the structural parameters reported in the preceding section relate to the physical behavior. We will first discuss the effect of stuffing on the AFM/MI transitions, and  then move on to the slope of $\rho$ versus T in the 'metallic' region.

\subsubsection {Specific heat and Magnetization}
\begin{figure}[!]
	\centering
	\includegraphics[width=0.5\textwidth]{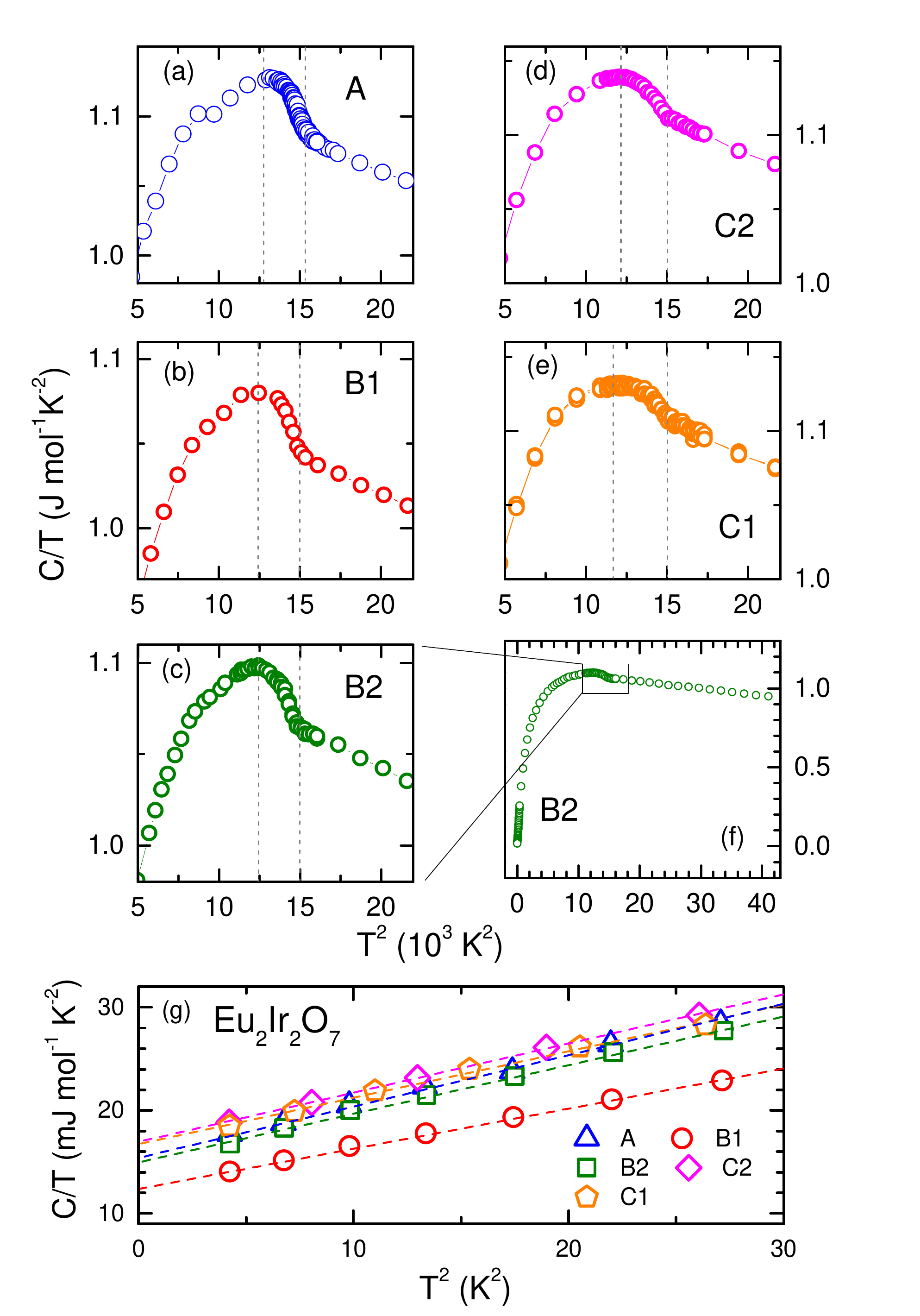}
	\caption{(a) to (e) Temperature (T) variation of specific heat (C$_P$) plotted as C$_p$/T  Vs. T$ ^2 $ for samples: A0, B1, B2, C2 \& C1 shown over a narrow temperature range around the AFM/MI transition. The vertical dashed lines are guide to eye to show the onset of transition and transition width; (f) C$_p$/T  Vs. T$ ^2 $ shown over the whole measurement temperature range for sample B2. (g) C$_p$/T  Vs. T$ ^2 $ for all the samples at low-temperatures. The dashed lines are straight-line fits to the data (see text for details)}
	\label{cbyt}
\end{figure}

Specific heat of samples A, B1, B2, C1 and C2, over a narrow temperature range near the magnetic transition is plotted as C$_p$/T  Vs. T$ ^2 $ in Fig. \ref{cbyt}(a - e). Being very close in compositions, C$_p$/T  of these samples nearly overlap, therefore, specific heat over the full temperature range is shown only for sample B2 as a representative case in panel f. In panels a to e, two vertical dashed lines are shown: the one at higher temperature (near 15000 K$ ^2 $) marks the onset of AFM ordering (T$_N$). The second line at lower temperature that passes through the maximum in C$_p$/T  is used to characterize the width of the magnetic transition. In each sample, C$_p$/T  near the maximum is close to 1.1 $ \pm $ 0.1 J mol$ ^{-1} $ K$ ^{-2} $. This value is in good agreement with previous reports\cite{Takatsu, Matsuhira2011}. The value of T$ _N $ (122 $\pm$ 1 K) is nearly the same for all the samples, except A for which it is slightly enhanced to 124 $\pm$ 1 K. The transition width increases with stuffing due to an increase in the structural disorder. For example, in samples C1 and C2 the transition has been considerably smeared out.

\label{spheat}

The low-temperature specific heat for all the samples is shown in panel g between T = 2 K and 5.5 K. At low-temperatures, only the long-wavelength acoustic phonons are excited and their contribution to specific heat varies as T$ ^3 $ for T $\ll$ the Debye temperature ($ \theta_D $). Fitting C$_p$  at low-temperatures to an expression of the form $ C_p  = \gamma T + \beta T^3 $ can, therefore, yield information concerning the electronic correlations. Since the AFM spin-wave contribution to C$_p$ in spin systems with a gapless spin-wave excitation spectrum also varies as T$ ^3 $ (see Ref. \citenum{Krankendonk}), it should not affect the determination of $\gamma$. In a spin system with a finite gap in the spin excitation spectrum, the specific heat at low temperatures is expected to decrease exponentially to zero, which is evidently not the case here, suggesting that the excitations are indeed gapless. C$_p$/T  data for each sample is satisfactorily fitted using the equation: C$_p/T = \gamma  + \beta T^2 $. As expected, $\beta$ does not vary much between these samples giving a value of nearly 0.77 $\pm$ 0.01 mJ mol$ ^{-1} $ K$ ^{-4} $. $ \gamma $, however, vary from $\sim$10 mJ mol$ ^{-1} $ K$ ^{-2} $ for B1 to $\sim$15 mJ mol$ ^{-1} $ K$ ^{-2} $ for C1 and C2. The average value is in good agreement with 13 mJ mol$ ^{-1} $ K$ ^{-2} $ reported previously\cite{Blacklock}. Measurements on B1 were repeated at a later time to confirm that $ \gamma $ for this sample is indeed the lowest. Assuming that only Ir's 5d electrons contribute to the linear term in C$_p$, the value of $ \gamma $ per Ir-mol ($\sim$6.5 mJ Ir-mol$ ^{-1} $ K$ ^{-2} $) is almost an order of magnitude higher than $ \gamma $ of Cu \cite{Kittel:ISSP} indicating moderately strong electronic correlations as predicted theoretically\cite{Krempa}. The important point to note is that $\gamma$ has a substantial sample dependence -it tends to increase upon Eu-stuffing. This increase is not necessarily a consequence of further enhancement of the electronic correlations due to stuffing; we believe that it might simply be an effect of the carrier doping as considered by Ishikawa et al. who reported $\gamma$ values as high as 26 mJ mol$^{-1}$K$^{-2}$ for their negatively stuffed, most conducting EIO sample \cite{Ishikawa}.

\begin{figure}[t]
	\centering
	\includegraphics[width=0.5\textwidth]{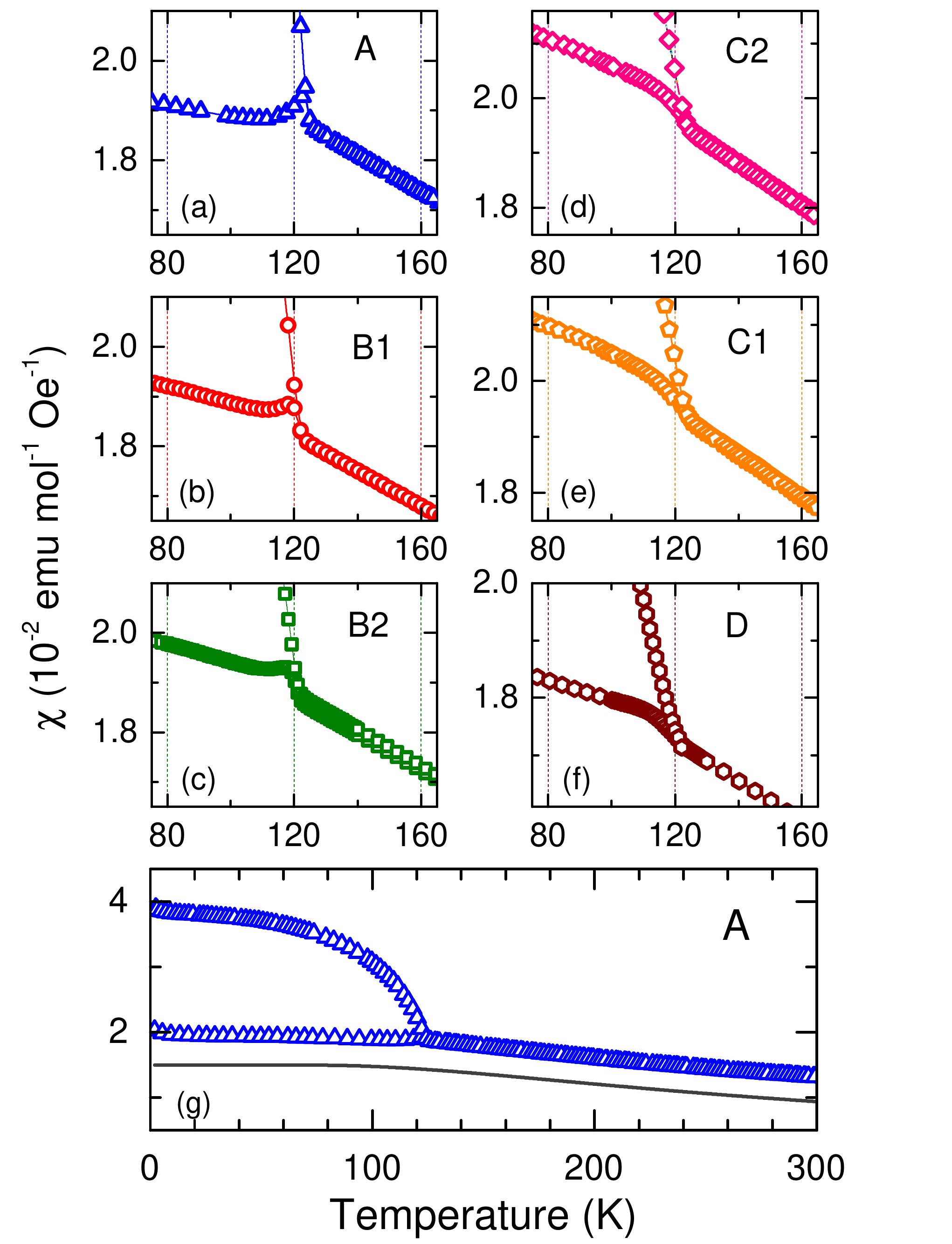}
	\caption {(a-f) Zero-Field-Cooled (ZFC) and Field-cooled (FC) susceptibility of all the Eu$_2$Ir$_2$O$_7$ samples is shown as a function of temperature in a small temperature range around the magnetic ordering temperature; (g) The ZFC and FC susceptibilities over the whole temperature range are shown for sample A as representative. $\chi_{VV}$ is van Vleck susceptibility of Eu$^{3+}$ ion (see text for details).} 
	\label{chiT}
\end{figure}

Temperature variation of magnetic susceptibility ($ \chi = M/H$) of samples A to D, measured under a static magnetic field of 1 kOe, is shown in Fig. \ref{chiT}(a-f). As a representative case, $\chi(T)$ of sample A is shown over the whole temperature range in panel g. When cooled below room temperature, $ \chi (T)$ of all the samples increases almost linearly down to T$_N \approx $ 124 K, where the Ir moments undergo long-range AFM ordering, which is shown to be of all-in/all-out (AIAO) type, wherein all \textit{four} Ir-moments on a given tetrahedron either point-in towards the center (AI) of the tetrahedron or point-out directly away from the center (AO) \cite{Donnerer, Sagayama}.  The data below T$_N$ show large ZFC-FC bifurcation. This bifurcation is believed to be due to a combination of several factors, including presence of 180$^\circ$ domain walls, anti-site disorder, Ir-vacancies and/or the occurrence of Ir$^{5+} $(see Ref. \citenum{Lef} for details).The value of magnetic transition temperature in our samples is in good agreement with similar values previously reported \cite{Matsuhira2011, Ishikawa}. Contribution of Ir moments to $\chi(T)$ can be estimated by subtracting the van Vleck term ($ \chi_{VV} $) due to Eu$^{3+}$. The calculated $ \chi_{VV} $ (taking the value of SO interaction $\lambda$ from Ref. \citenum{EuVanVleck}) is shown in Fig. \ref{chiT}. Below room temperature $ \chi_{VV} $ increases with almost the same slope as $\chi$(T), and below a temperature close to T$_N$ it tends to saturate. Thus, , above T$ _N $, $\chi$(T) owes its temperature dependence almost entirely to $ \chi_{VV} $. In another words, contribution of Ir moments to $\chi(T)$ appears to be almost temperature independent above T$ _N $. This behavior suggest that the Ir \textit{5d} electrons are itinerant and undergo a partial localization when cooled below T$ _N $, which qualitatively agrees with the experimental fact that resistivity (next section) also increases sharply below this temperature.      

We now examine how the transition temperature in $\chi$ is affected due to stuffing in our samples. In A, B1 and B2, the ordering is marked by the presence of a cusp in $ \chi(T) $ below which the ZFC and FC bifurcates out. In C1, C2 and D no cusp is seen, the transition has rather smeared out, in agreement with the specific heat. For sample A, T$ _N $ (position of the cusp) appears to be slightly higher compared to the other samples. 

In short, specific heat and susceptibility data reveal the following information: (i) no significant change in the value of T$_N$, (ii) broadening of the transition width upon stuffing (or increasing disorder), (iii) the shape of anomaly associated with AFM ordering in both C$_p$  and $\chi$ are nearly identical for samples prepared in a given batch, and (iv) a closer look at the magnitude of  $\chi$ at any fixed temperature (say, at T = 80 K) reveals that the value of $\chi$ scales approximately with the ratio Eu : Ir in our samples. For example, $\chi$ of sample D: $\sim$0.00184 emu mol$ ^{-1} $Oe$ ^{-1} $  $<$  A \& B1: $\sim$0.0192 emu mol$ ^{-1} $Oe$ ^{-1} $ $<$ B2: 0.0198 emu mol$ ^{-1} $Oe$ ^{-1} $ $ <$ C1 \& C2: $\sim$0.0212 emu mol$ ^{-1} $Oe$ ^{-1} $. This scaling is expected since the concentration of Eu$^{3+} $ determines $ \chi_{VV} $ which has a significant contribution to the total measured $\chi$ as depicted in Fig. \ref{chiT} (lowest panel).

\subsubsection{Resistivity and thermoelectric power}
\label{resistivity}
\begin{figure}[!]
	\centering
	\includegraphics[width=0.5\textwidth]{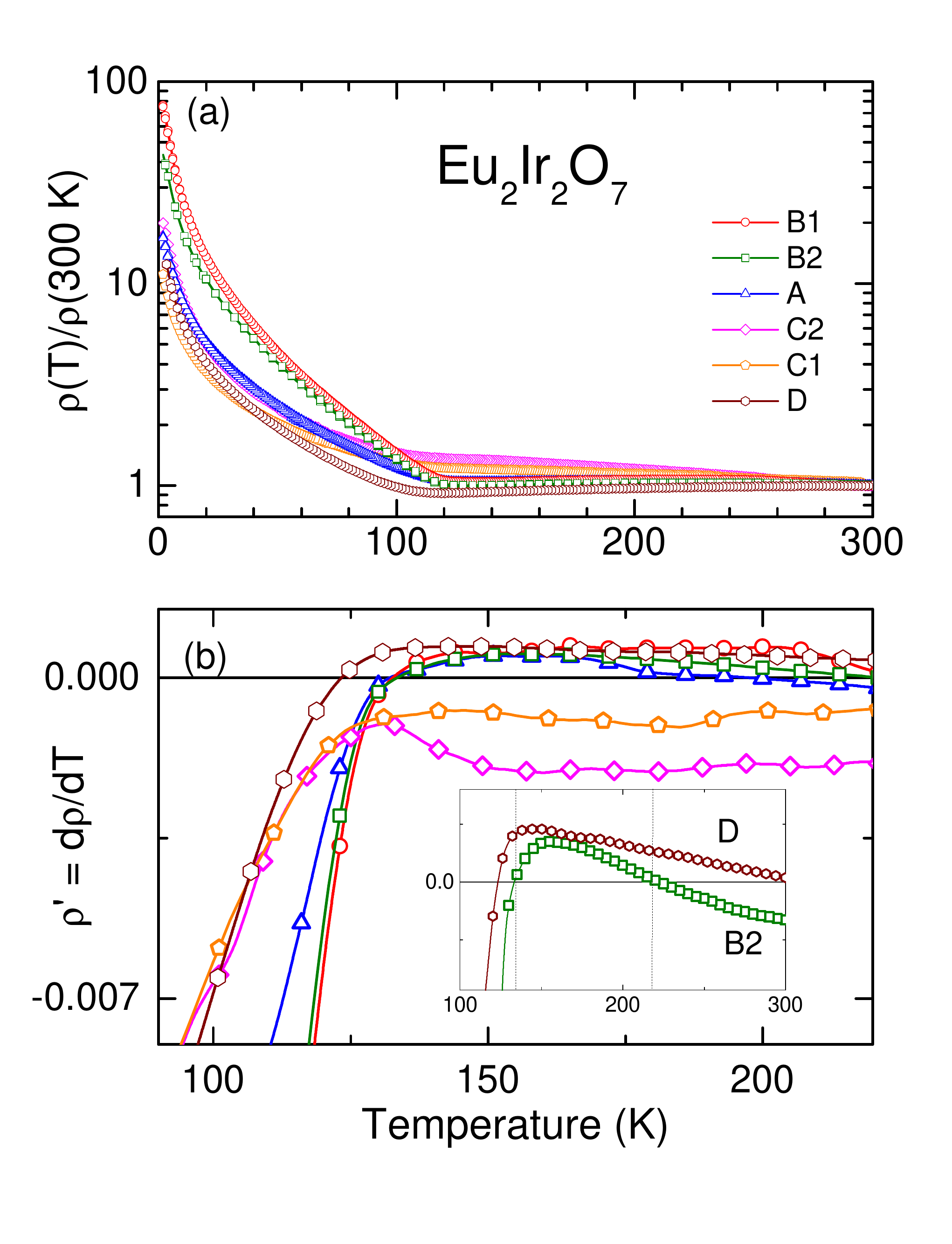}
	\caption {(a) $\rho(T)/\rho(300K)$ plotted as a function of temperature (T) for various Eu$ _2 $Ir$ _2 $O$ _7 $ samples (A, B1, B2, C2, C1 \& D); (b) first derivative  $\rho ' = d\rho/dT$ of the resistivity plots as a function of temperature around the metal-insulator transition. In the inset $\rho '$ is shown for samples B2 and D over a wider temperature range to demonstrate the change of sign of $d\rho/dT$.}  
	\label{rho}
\end{figure}

In Fig. \ref{rho}a, normalized resistivity ($\rho_N =  \rho (T)/ \rho (300 K)$) is shown as a function of temperature for all the EIO samples, where $\rho(300 K)$ is the value of resistivity at T = 300 K. $\rho_N$ in each case exhibits a sharp increase upon cooling below  T$ _\textit{MI} \approx $ 125 K. The MI transition temperature is estimated from the derivative plots by linearly extrapolating $ d\rho/dT $ data above and below the MI transition. The temperature where these lines intersects is taken as T$ _\textit{MI} $. Using this criteria, T$_\textit{MI} $ for samples A, B1 and B2 is nearly 127 K, for samples C1 and C2 it is close to 122 K, and for sample D it is around 120 K. The average value of T$ _\textit{MI} $ is 124 K, which shows a good agreement with the previous reports \cite{Matsuhira2007, Tafti}, and with the values of T$_N$ obtained from C$_p$(T) and $\chi$(T) data presented in the preceding section. However, the temperature T$_\textit{MI}$ suppresses in the stuffed samples while T$_N$ remains nearly constant, which suggests that in a sample with even higher Eu-Ir off-stoichiometry, these temperatures may further separate out, as in the case of Nd$ _2 $Ir$ _2 $O$ _7 $, where such a separation has been previously reported \cite{Graf}.  

We now focus on the differences in $\rho(T)$ behavior of various EIO samples arising due to off-stoichiometry. An important difference concerns the sign of $\rho ' = d\rho/dT$ above T$_\textit{MI} $. For samples A, B1 and B2, d$\rho$/dT changes sign from negative to positive upon cooling below a certain temperature T$^*$ that lies above T$_\textit{MI}$ as shown in the inset of Fig. \ref{rho}b. T$ ^* $ is $\sim$ 170 $\pm$ 20 K for A and close to 230 $\pm$ 10 K for B1 and B2; and for sample D the slope is positive but it is about to become zero as temperature approaches 300 K, indicating that even for this sample a sign change is expected at higher temperatures above 300 K. On the other hand, for samples C1 and C2, the sign of d$\rho$/dT remains negative at all temperatures above T$ _\textit{MI} $.  

Briefly, two different d$\rho/$dT behaviors above T$_\textit{MI} $ are observed: samples with larger lattice parameter and higher Eu-stuffing (C1 and C2) exhibit a negative d$\rho/$dT for all temperatures above T$_{MI}$; samples with smaller lattice parameter and Eu/Ir ratio close to 1 (A, B1, B2 and D) exhibit a change of sign of d$\rho/$dT above T$_{MI}$. This behavior is analogous to that under external pressure investigated by Tafti et al. \cite{Tafti}. They show that as the applied pressure increases the sign of d$\rho/$dT changes from negative to positive, with an intermediate range of pressures where d$\rho/$dT changes sign at some temperature T$ ^* $ $>$ T$_\textit{MI}$. This analogy is not surprising since the lattice parameter itself varies with pressure. Thus, in both studies its the lattice parameter variation that drives the change of sign of d$\rho$/dT.    

\begin{figure}[hbtp]
	\centering
	\includegraphics[width=0.49\textwidth]{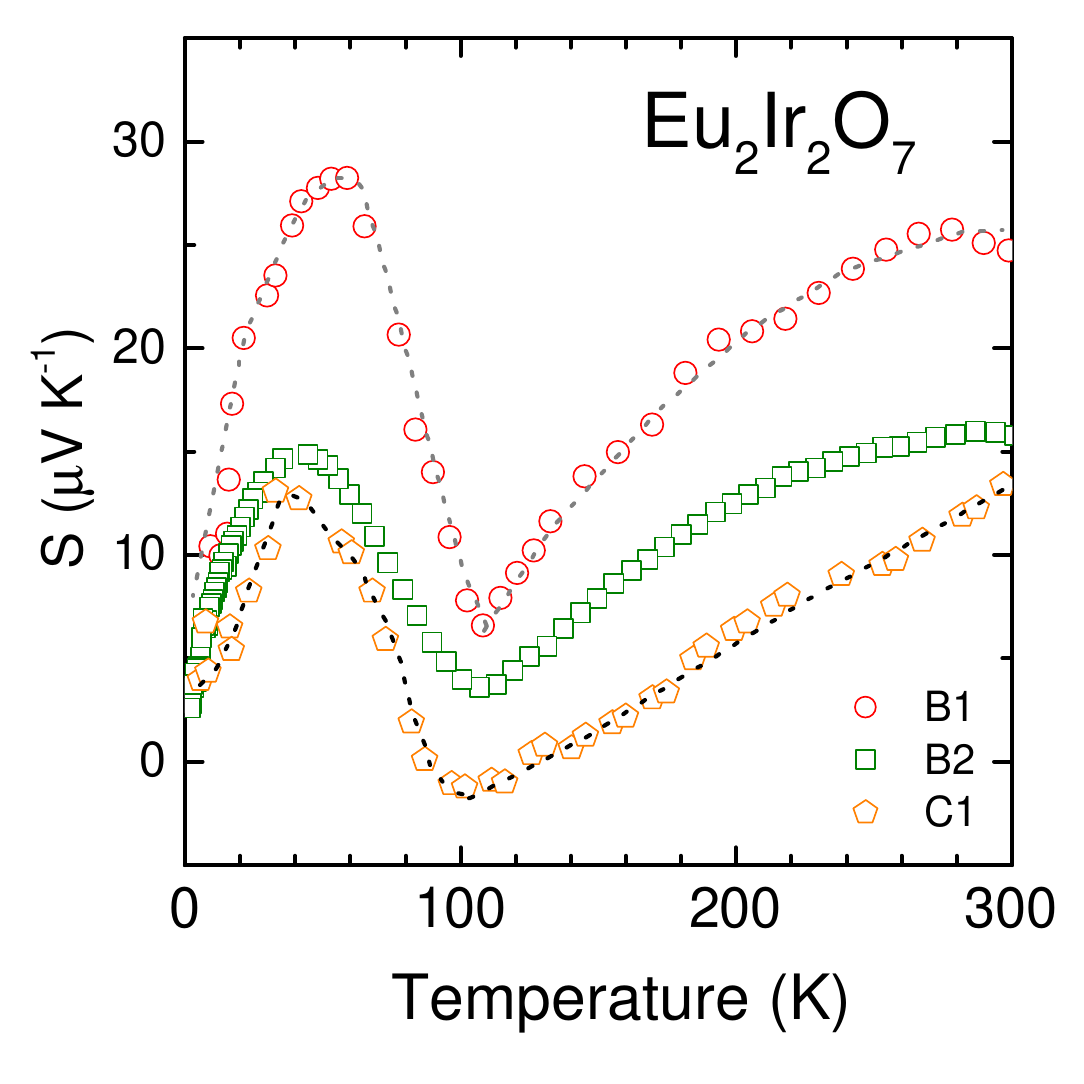}
	\caption {Thermoelectric power (S) plotted as a function of temperature for various Eu$ _2 $Ir$ _2 $O$ _7 $ samples: B1, B2 \& C1. The dotted lines are a guide to the eye.}  
	\label{S}
\end{figure}

While the sign of d$\rho$/dT in our samples and under applied pressure is in complete agreement, the magnitude of $\rho$ differs in the two cases. In the pressure study, not only the sign of d$\rho$/dT changed continuously from negative to positive with increasing pressure, the magnitude of $\rho$ also decreased simultaneously. In the present study, on the other hand, $\rho$ of sample C1 (decompressed analog of pressure study) having a negative d$\rho$/dT is 100 m$\Omega$ cm at 2 K which is smaller than the corresponding value for sample A1 ($\sim$7000 m$\Omega$ cm) which has a smaller lattice parameter compared to C1. Sample D, on the other hand, exhibits both a smaller lattice parameter and a smaller resistivity, comparable to that of C1. This, rather uncorrelated, variation of $\rho$ in our samples may arise from the fact that the magnitude of resistivity in a sintered pellet, unlike a single crystal specimen under pressure, depends on several factors including, the presence or absence of specific impurity phases that tend to accumulate along the grain boundaries and affect the electrical conduction, carrier doping due to stuffing, microstructure of the pellets and the oxygen vacancies, if any. 

Due to similar conditions used in the preparation of final sintered pellets employed for physical properties measurements in each case, we do not expect either the microstructure or the oxygen vacancies to vary very significantly in our samples. On the other hand, the level of Eu-stuffing, which differs in our samples, and the presence of metallic impurities may have affected the magnitude of $\rho$. To understand how stuffing leads to carrier doping we need to consider the charge difference between Eu$^{3+}$ and Ir$^{4+}$. If a sample is synthesized under an inert atmosphere, i.e., if in the surrounding medium there is no oxygen for the sample to absorb then the final product will form with oxygen vacancies. It will therefore have a chemical formula of the form: Eu$_{2+s}$Ir$_{2-s}$O$_{7-s/2}$, obtained using the charge-neutrality condition. However, when synthesized in air or under an oxygen flow, due to oxygen absorption during the synthesis, the sample is expected to form with fewer oxygen vacancies than before, and its chemical formula will be: Eu$_{2+s}$Ir$_{2-s}$O$_{7 + \delta - s/2}$, where $\delta$ is a non-negative quantity that corresponds to the hole doping in the sample. In the stuffed samples, therefore, the magnitude of $\rho$ can decrease considerably due to these additional charge carriers. 

With this insight, we now compare the resistivity of sample B1 to that of C1. We notice a significant decrease in $ \rho $, approximately 7000 m$\Omega$ cm (B1) to $\sim$100 m$\Omega$ cm (C1), which we believe is mainly because of the carrier doping in C1 as a consequence of stuffing. On the other hand, a comparison of samples prepared in the same batch revealed a much smaller difference (for example, $\sim$100 m$\Omega cm $ at 2 K for C1 and $\sim$750 m$\Omega cm $ at the same temperature for C2), which is probably due the presence of conducting IrO$_2$ as a parasitic phase in sample C1, and also a slightly higher Eu stuffing level for this sample. The resistivity of sample D is considerably reduced compared to that of B1 or B2, even though they have only minor differences in their lattice parameter. This could be due to the presence of Ir-metal as a parasitic phase in sample D, and probably the electron doping due to slight negative stuffing is also contributing to it.    

We examined the oxygen content in samples C1 and C2 using the thermogravimetric analysis. This was done under flowing O$_2$ up to a temperature of 1100 $^\circ$C. In these experiments, however, we failed to detect any weight gaining step up to the highest temperature, which suggests that there are few, if any, oxygen vacancies in these samples, i.e., the sample forms with a stoichiometry Eu$_{2+s}$Ir$_{2-s}$O$_{7}$. If that is the case then the Eu-stuffed samples will be hole doped. However, attempts to measure the carrier concentration directly using the Hall effect by applying fields up to $\pm$ 3 Tesla on the sintered pellets turned out to be unsuccessful. In our Hall measurements, which we did in the five-probe geometry using both positive and negative fields to eliminate the longitudinal voltage drop, we found irreproducible hysteretic behavior. 

To mitigate this difficulty to some extent, we carried out thermopower measurements on some of our samples. Thermopower, denoted by S, is a complex physical quantity due to its dependence on various factors \cite{Macdonald}, including the electronic diffusion under temperature gradient which produces a thermopower proportional to T, and the electron-phonon interaction resulting in the phonon drag contribution. Thermopower of our samples A, B2 and C1 is shown as a function of temperature in Fig.\ref{S}. Upon cooling, S decreases monotonically down to T $ \approx$ 120 K, which is the expected behavior for metals at high temperatures. Below 120 K, however, it increases rather sharply. This change of behavior near 120 K is most likely related to the metal-insulator transition. Upon further lowering the temperature, S exhibits a peak around T$_\textit{peak}$ = 50 K. This peak is likely a manifestation of  phonon-drag  which increases S as T$^3$ at low-temperatures and decays as 1/T at higher temperatures, resulting in a broad peak in the intermediate temperature range\cite{Macdonald}. T$_\textit{peak}$ appears to have a slight sample dependence. The sign of S remains positive over the entire temperature range, which indicates that the majority carriers in EIO are holes. The overall S(T) behavior of our samples is in a fairly good agreement with the previous reports \cite{Matsuhira2007, Matsuhira2011}. A comparison of the thermopower of several EIO samples, including those from the present study, show that the qualitative features in the temperature variation of thermopower are more or less sample independent, but the magnitude of thermopower scales directly with sample's resistivity. For example, B1 which is the most resistive sample in our study exhibits the highest thermopower, and C1 which has the smallest resistivity also has the smallest thermopower. A review of the previous literature revealed a similar trend in the thermopower of EIO samples investigated by various groups. Bouchard et al. \cite{BOUCHARD1971669}, for example, reported a thermopower of approximately 10$\mu$V/K at 300 K for their air-synthesized sample ($\rho$ (300 K) $\approx$ 20 $m\Omega$ cm). The corresponding value for a vacuum synthesized sample, reported by Matsuhira et al. \cite{Matsuhira2011} is about 40 $\mu$V/K ($\rho$ (300 K) $\approx$ 100 $m\Omega$ cm) \cite{Matsuhira2011}. As a passing remark, we note the presence of a small curvature in the thermopowers of samples B1 and B2 in the temperature range 120 K $<$ T $<$ 300 K; this may be related to the sign change of d$\rho$/dT for these samples. In C1 no such curvature is seen. 

To summarize, thermopower, which varies as inverse of the carrier concentration in a simple one band model\cite{Macdonald}, also supports the conclusion that stuffing leads to hole doping which simultaneously decreases the thermopower and the resistivity, and increases the coefficient of linear term in the specific heat. 
  
\section{Discussion on the sign of $d\rho/dT$}
\label{disc} 

The structural refinements carried out on very high-resolution synchrotron X-ray diffraction data collected for all the EIO samples synthesized in this work led to a precise determination of the lattice parameter, and an estimate of the Eu : Ir stoichiometry of the main pyrochlore phase (table \ref{Table1}). As depicted in Fig. \ref{peakshift}, where the position of 622 Bragg peak is shown as a reference, the lattice parameter of our samples varies in the following order: $a_{C1} > a_{C2} > a_{B2} > a_{B1} >  a_A > a_D$. This variation can be understood on the basis of Eu$^{3+}$-stuffing at the Ir-site or vice-versa as has been extensively cited for stuffed titanate pyrochlores \cite{Lau, Rossstuffedtitanates}. 

All the EIO samples investigated here showed a weakly temperature dependent resistivity in the metallic state above T$_\textit{MI}$, which is the hallmark of EIO amongst the entire iridate pyrochlore series. However, the sign of d$\rho$/dT in the metallic state showed a systematic sample dependence. It changed from negative for samples C1 \& C2 having a relatively larger lattice parameters to positive for sample D having the smallest lattice parameter. For samples A, B1 \& B2, d$\rho$/dT changed its sign at an intermediate temperature T$^*$, such that, between T$_\textit{MI}$ and T$^*$, d$\rho$/dT remains positive as it should be for a metal; and for T $>$ T$^*$, it becomes negative. Even for sample D, d$\rho$/dT approaches zero as T goes to 300 K, indicating that even for this sample a change of sign is expected at higher temperatures. It should be noted that while the sign of d$\rho$/dT is dependent on the lattice parameter, the temperature T$_\textit{MI}$ does not, which remains nearly unchanged for our samples. The maximum overall change in T$_\textit{MI}$ across our samples remained less than 5 \%. 

Interestingly, the behavior of d$\rho$/dT with lattice parameters or equivalently with the chemical pressure induced by stuffing mimics exactly the behavior under externally applied pressure investigated by Tafti et al. \cite{Tafti}. In the pressure study, Tafti et al. used an EIO sample having a negative d$\rho$/dT in the 'metallic' ragime analogous to samples C1 or C2 in our study. They showed that d$\rho$/dT remained negative under pressures up to about 4.5 GPa, and the sign changed to positive when the pressure exceeded 7 GPa or so. In the intermediate range of pressures, d$\rho$/dT changed it sign at a temperature T$^*$ $>$ T$_\textit{MI}$ in an exact analogy with our results. In the pressure experiments also, while d$\rho$/dT changed continuously with increasing pressure, T$_\textit{MI}$ shifted only marginally. Tafti et al. called the 'metallic' regime above T$_\textit{MI}$ with negative d$\rho$/dT as an "incoherent" metal as opposed to the conventional metallic behavior where the sign of d$\rho$/dT is expected to remain positive. 

Our preliminary X-ray diffraction study under high-pressure at 300 K on sample A at ELETTRA shows that indeed a lattice contraction  comparable to the difference between the lattice parameters of samples C1 and D can easily be produced under moderate pressures ($\sim$0.01 \AA, from 4 to 7 GPa-the range over which d$\rho$/dT changes is sign from negative to positive in the pressure study\cite{Tafti}). From this observation it can be concluded that the change in lattice parameter due to a minor, and often inadvertent, stuffing resulting from a loss of IrO$_2$ during sintering, can indeed be comparable to that under moderate pressure.

Since increasing the level of Eu-stuffing enhances the structural disorder, one may argue that the incoherent behavior is likely a consequence of the increased disorder which results in a weak-localization of the charge carriers. However, we note that in the pyrochlore structure, the increase in applied pressure has a tendency to enhance the antisite disorder (see for example, ref. \citenum{Minervini} and references therein). Thus, if disorder induced weak-localization is indeed leading to the incoherent behavior then we expect d$\rho$/dT to remain negative under pressure, which is contrary to the experimental observation where the increase in pressure enhances the metallic behavior. It should also be pointed out that in the KF-flux grown Ir-rich EIO single crystals\cite{Ishikawa}, d$\rho$/dT remains negative up to the highest level of Ir stuffing. Since these samples are structurally disordered due to Ir stuffing, their metal-like resistivity suggest that disorder cannot be the primary reason for the incoherent-metal behavior. In a recent theoretical study, it has been argued that the effect of small disorder in 3D semimetals with intermediate to strong interaction can in general be disregarded due to an effective screening of the disorder that raises the critical disorder strength for disorder-driven diffusive transition to a higher threshold \cite{Gomzalez}. Thus, we have very little or no reason to believe that the sign of d$\rho$/dT is controlled by the level of disorder in the sample.         

We now discuss the structural changes taking place upon changing the lattice parameter to understand the anomalous sign of d$\rho$/dT in the metallic regime. It has been highlighted in several previous studies (see for example, Refs. \citenum {Ishii} \& \citenum {HongbinZhang}) that the Ir$-$O$-$Ir bond angle is an important parameter that controls the $t_\textit{2g}$ bandwidth, and the strength of the magnetic superexchange interaction between the Ir-moments. Previous pressure studies on some other pyrochlores suggest that \textit{u} changes only marginally in the presence of moderate pressures \cite{Apetrei, Saha, Clancy}. Since the Ir$-$O$-$Ir bond angle depends on \textit{u}, it is also expected to remain almost invariant under pressure. Furthermore, since T$_N$/T$_\textit{MI}$ in our samples has not changed significantly, we believe that the changes in the Ir$-$O$-$Ir bond angle due to stuffing is not expected to be very significant. This suggests that either d$\rho$/dT in EIO is extremely sensitive to the Ir$-$O$-$Ir bond angle, or, perhaps, the direct Ir$-$Ir hopping, which is expected to be significant due to the extended nature of Iridium's 5d orbitals, also plays a role here. Unlike the Ir$-$O$-$Ir bond angle, the dependence of direct hopping on the lattice parameter is straightforward: as the lattice contracts the Ir$-$Ir distance decreases, increasing the direct overlap between the 5d orbitals.    

The role of direct hopping in the presence of indirect Ir$-$O$-$Ir hopping was considered in the theoretical study by Witczak-Krempa et al.\cite{Krempa2012}. In Fig. 2 of their paper, it is shown that for a fixed Hubbard U, and for a given value of the transfer integral $t_\textit{Ir-O-Ir}$ corresponding to the Ir$-$O$-$Ir hopping, various phases, including metal, topological semimetal (TSM), topological insulator(TI), and topologically trivial insulators can be stabilized by tuning the Ir$-$Ir distance in the proximity of the metal-TSM-insulator phase boundary. In particular, the TSM phase is shown to appear only over a very narrow range of values of $t_\textit{Ir-Ir}$, i.e., the transfer integral corresponding to the Ir-Ir hopping. Thus, with a small variation of Ir$-$Ir distance either a metallic or an insulating phase can be stabilized. On the other hand, away from the metal-TSM-insulator phase boundary, direct hopping becomes less relevant, which is probably the reason why in Nd$_2$Ir$_2$O$_7$ (NIO), which is located somewhat away from this boundary, the sign of d$\rho$/dT remains unchanged (positive) under pressure \cite{Sakata}. 
  
EIO, on the other hand, is located in a close proximity of the metal-TSM-insulator phase boundary and, therefore, even minor changes in the lattice parameter, either due to applied pressure or due to stuffing, is expected to have a significant effect on the electrical transport behavior, probably not as much due to the Ir$-$O$-$Ir bond angle as to the Ir$-$Ir bond distance. If this is indeed is true then for the physical realization of TSM phase in EIO, which purportedly harbors the Weyl fermions, it is critical to optimize the Eu : Ir ratio to get the optimal Ir$-$Ir distance and carrier density. This might be the reason why till date there are only few, only one to the best of our knowledge, experimental evidences of the Weyl semimetallic ground state in EIO. 

 \section{Summary \& conclusions}
\label{conc}

Briefly, this work aimed at understanding the sample dependence of electrical behavior of the pyrochlore EIO which, along with the neighboring members of this family, has been predicted to show interesting topological phases. In the past literature, two types of resistivity behavior were reported for EIO, one where d$\rho$/dT  remains negative above T$_\textit{MI}$  (type I), and second where it remains positive (type II). Alongside, a pressure study on an EIO sample of type I (d$\rho$/dT $<$ 0) revealed: (i) d$\rho$/dT $<$ 0 for pressures up to 4.6 GPa, (ii) d$\rho$/dT  changing sign at a temperature T$^*$ ($>$ T$_\textit{MI}$) for pressures in the range ~ 4.6 to ~ 7 GPa, and (iii) d$\rho$/dT $>$ 0 for pressures exceeding 7 GPa \cite{Tafti}.
In this study, we prepared six different EIO samples with slightly varying compositions to reproduce all the three d$\rho$/dT behaviors ((i), (ii) and (iii)) listed above at ambient pressure. In order to establish a correspondence between the d$\rho$/dT  behaviors in our various EIO samples with that under pressure in ref. \citenum{Tafti}, we did high-resolution synchrotron powder X-ray diffraction at the MCX beamline in ELETTRA on all the samples. These experiments successfully revealed minor variations in the unit cell volume of our samples which established that it is the chemical pressure that imitates the external pressure. The Rietveld refinement of the X-ray data further revealed that the observed variation in the unit cell volume is due to Eu-stuffing at the Ir-site. This not only allowed shedding light on the peculiar d$\rho$/dT behavior under pressure but it also resolved the enigmatic sample dependence of d$\rho$/dT in the previous reports. Eu-stuffing in EIO results from loss of volatile IrO$_2$ from the reaction mixture during high-temperature sintering. The experimental results obtained on the various EIO samples suggest that the properties of EIO are not as much sensitive to small variation in the starting composition as they are to the synthesis protocol. This sensitivity is a consequence of the loss of volatile IrO$_2$ during the sintering as the temperature exceeds 1050 $^\circ$C or so. Our results indicate an extreme sensitivity of EIO to pressure: applied, or chemical due to slight Eu-stuffing. We believe this extreme sensitivity is a due to the close proximity of EIO to the metal-insulator phase boundary, which makes the Ir-Ir hopping an important parameter in deciding the ground state of EIO. In future, it will be useful to investigate stuffing in other pyrochlore iridates (e.g., Sm$_2$Ir$_2$O$_7$) that also lie close to the metal-insulator phase boundary.

\section*{Acknowledgments}
The authors acknowledges financial support from DST/SERB India under grant nos. EMR/2016/003792/PHY and SR/NM/TP-13/2016. PT and SS thank DST for financial support to perform experiments at Elettra,Italy. AKS thanks Department of Science and Technology, India for financial support. The authors would like to thank Dr. Lara Gigli for all the help during experiments at MCX beamline, Elettra, Italy.

\section*{References}   
\nocite{*}
\bibliography{Iridates}

\begin{thebibliography}{10}
\providecommand*{\bibinfo}[2]{#2}
\providecommand*{\eprint}[1]{#1}
\providecommand*{\url}[1]{#1}
\bibitem{Gardner}
\bibinfo{author}{J.~S. Gardner}, \bibinfo{author}{M.~J.~P. Gingras}, and
  \bibinfo{author}{J.~E. Greedan}, \bibinfo{journal}{Rev. Mod. Phys.}
  \bibinfo{volume}{\textbf{82}}, \bibinfo{pages}{53} (\bibinfo{date}{Jan
  2010}), \url{http://link.aps.org/doi/10.1103/RevModPhys.82.53}.
\bibitem{Balents}
\bibinfo{author}{L.~Balents}, \bibinfo{journal}{Nature}
  \bibinfo{volume}{\textbf{464}}, \bibinfo{pages}{199 } (\bibinfo{date}{2010}),
  \url{http://dx.doi.org/10.1038/nature08917}.
\bibitem{WanWeyl}
\bibinfo{author}{X.~Wan}, \bibinfo{author}{A.~M. Turner},
  \bibinfo{author}{A.~Vishwanath}, and \bibinfo{author}{S.~Y. Savrasov},
  \bibinfo{journal}{Phys. Rev. B} \bibinfo{volume}{\textbf{83}},
  \bibinfo{pages}{205101} (\bibinfo{date}{May 2011}),
  \url{http://link.aps.org/doi/10.1103/PhysRevB.83.205101}.
\bibitem{Krempa}
\bibinfo{author}{W.~Witczak-Krempa}, \bibinfo{author}{A.~Go}, and
  \bibinfo{author}{Y.~B. Kim}, \bibinfo{journal}{Phys. Rev. B}
  \bibinfo{volume}{\textbf{87}}, \bibinfo{pages}{155101} (\bibinfo{date}{Apr
  2013}), \url{http://link.aps.org/doi/10.1103/PhysRevB.87.155101}.
\bibitem{Hongbin}
\bibinfo{author}{H.~Zhang}, \bibinfo{author}{K.~Haule}, and
  \bibinfo{author}{D.~Vanderbilt}, \bibinfo{journal}{Phys. Rev. Lett.}
  \bibinfo{volume}{\textbf{118}}, \bibinfo{pages}{026404} (\bibinfo{date}{Jan
  2017}), \url{https://link.aps.org/doi/10.1103/PhysRevLett.118.026404}.
\bibitem{Millis}
\bibinfo{author}{R.~Wang}, \bibinfo{author}{A.~Go}, and \bibinfo{author}{A.~J.
  Millis}, \bibinfo{journal}{Phys. Rev. B} \bibinfo{volume}{\textbf{95}},
  \bibinfo{pages}{045133} (\bibinfo{date}{Jan 2017}),
  \url{https://link.aps.org/doi/10.1103/PhysRevB.95.045133}.
\bibitem{Tokiwa}
\bibinfo{author}{Y.~Tokiwa}, \bibinfo{author}{J.~J. Ishikawa},
  \bibinfo{author}{S.~Nakatsuji}, and \bibinfo{author}{P.~Gegenwart},
  \bibinfo{journal}{Nature Materials} \bibinfo{volume}{\textbf{13}},
  \bibinfo{pages}{356} (\bibinfo{date}{2014}),
  \url{http://dx.doi.org/10.1038/nmat3900 L3 - 10.1038/nmat3900}.
\bibitem{Matsuhira2011}
\bibinfo{author}{K.~Matsuhira}, \bibinfo{author}{M.~Wakeshima},
  \bibinfo{author}{Y.~Hinatsu}, and \bibinfo{author}{S.~Takagi},
  \bibinfo{journal}{Journal of the Physical Society of Japan}
  \bibinfo{volume}{\textbf{80}}(9), \bibinfo{pages}{094701}
  (\bibinfo{date}{2011}), \url{http://dx.doi.org/10.1143/JPSJ.80.094701}.
\bibitem{shushkov}
\bibinfo{author}{A.~B. Sushkov}, \bibinfo{author}{J.~B. Hofmann},
  \bibinfo{author}{G.~S. Jenkins}, \bibinfo{author}{J.~Ishikawa},
  \bibinfo{author}{S.~Nakatsuji}, \bibinfo{author}{S.~Das~Sarma}, and
  \bibinfo{author}{H.~D. Drew}, \bibinfo{journal}{Phys. Rev. B}
  \bibinfo{volume}{\textbf{92}}, \bibinfo{pages}{241108} (\bibinfo{date}{Dec
  2015}), \url{https://link.aps.org/doi/10.1103/PhysRevB.92.241108}.
\bibitem{Zhao}
\bibinfo{author}{S.~Zhao}, \bibinfo{author}{J.~M. Mackie},
  \bibinfo{author}{D.~E. MacLaughlin}, \bibinfo{author}{O.~O. Bernal},
  \bibinfo{author}{J.~J. Ishikawa}, \bibinfo{author}{Y.~Ohta}, and
  \bibinfo{author}{S.~Nakatsuji}, \bibinfo{journal}{Phys. Rev. B}
  \bibinfo{volume}{\textbf{83}}, \bibinfo{pages}{180402} (\bibinfo{date}{May
  2011}), \url{https://link.aps.org/doi/10.1103/PhysRevB.83.180402}.
\bibitem{Takatsu}
\bibinfo{author}{H.~Takatsu}, \bibinfo{author}{K.~Watanabe},
  \bibinfo{author}{K.~Goto}, and \bibinfo{author}{H.~Kadowaki},
  \bibinfo{journal}{Phys. Rev. B} \bibinfo{volume}{\textbf{90}},
  \bibinfo{pages}{235110} (\bibinfo{date}{Dec 2014}),
  \url{https://link.aps.org/doi/10.1103/PhysRevB.90.235110}.
\bibitem{Matsuhira2007}
\bibinfo{author}{K.~Matsuhira}, \bibinfo{author}{M.~Wakeshima},
  \bibinfo{author}{R.~Nakanishi}, \bibinfo{author}{T.~Yamada},
  \bibinfo{author}{A.~Nakamura}, \bibinfo{author}{W.~Kawano},
  \bibinfo{author}{S.~Takagi}, and \bibinfo{author}{Y.~Hinatsu},
  \bibinfo{journal}{Journal of the Physical Society of Japan}
  \bibinfo{volume}{\textbf{76}}(4), \bibinfo{pages}{043706}
  (\bibinfo{date}{2007}), \eprint{http://dx.doi.org/10.1143/JPSJ.76.043706},
  \url{http://dx.doi.org/10.1143/JPSJ.76.043706}.
\bibitem{Ishikawa}
\bibinfo{author}{J.~J. Ishikawa}, \bibinfo{author}{E.~C.~T. O'Farrell}, and
  \bibinfo{author}{S.~Nakatsuji}, \bibinfo{journal}{Phys. Rev. B}
  \bibinfo{volume}{\textbf{85}}, \bibinfo{pages}{245109} (\bibinfo{date}{Jun
  2012}), \url{http://link.aps.org/doi/10.1103/PhysRevB.85.245109}.
\bibitem{Clancy}
\bibinfo{author}{J.~P. Clancy}, \bibinfo{author}{H.~Gretarsson},
  \bibinfo{author}{E.~K.~H. Lee}, \bibinfo{author}{D.~Tian},
  \bibinfo{author}{J.~Kim}, \bibinfo{author}{M.~H. Upton},
  \bibinfo{author}{D.~Casa}, \bibinfo{author}{T.~Gog},
  \bibinfo{author}{Z.~Islam}, \bibinfo{author}{B.-G. Jeon}, \emph{et~al.},
  \bibinfo{journal}{Physical Review B} \bibinfo{volume}{\textbf{94}}(2)
  (\bibinfo{date}{2016}), \url{http://dx.doi.org/10.1103/PhysRevB.94.024408}.
\bibitem{Tafti}
\bibinfo{author}{F.~F. Tafti}, \bibinfo{author}{J.~J. Ishikawa},
  \bibinfo{author}{A.~McCollam}, \bibinfo{author}{S.~Nakatsuji}, and
  \bibinfo{author}{S.~R. Julian}, \bibinfo{journal}{Phys. Rev. B}
  \bibinfo{volume}{\textbf{85}}, \bibinfo{pages}{205104} (\bibinfo{date}{May
  2012}), \url{http://link.aps.org/doi/10.1103/PhysRevB.85.205104}.
\bibitem{Gd2Ti2O7}
\bibinfo{author}{P.~Nachimuthu}, \bibinfo{author}{S.~Thevuthasan},
  \bibinfo{author}{M.~H. Engelhard}, \bibinfo{author}{W.~J. Weber},
  \bibinfo{author}{D.~K. Shuh}, \bibinfo{author}{N.~M. Hamdan},
  \bibinfo{author}{B.~S. Mun}, \bibinfo{author}{E.~M. Adams},
  \bibinfo{author}{D.~E. McCready}, \bibinfo{author}{V.~Shutthanandan},
  \emph{et~al.}, \bibinfo{journal}{Phys. Rev. B} \bibinfo{volume}{\textbf{70}},
  \bibinfo{pages}{100101} (\bibinfo{date}{Sep 2004}),
  \url{https://link.aps.org/doi/10.1103/PhysRevB.70.100101}.
\bibitem{Baroudi}
\bibinfo{author}{K.~Baroudi}, \bibinfo{author}{B.~D. Gaulin},
  \bibinfo{author}{S.~H. Lapidus}, \bibinfo{author}{J.~Gaudet}, and
  \bibinfo{author}{R.~J. Cava}, \bibinfo{journal}{Phys. Rev. B}
  \bibinfo{volume}{\textbf{92}}, \bibinfo{pages}{024110} (\bibinfo{date}{Jul
  2015}), \url{http://link.aps.org/doi/10.1103/PhysRevB.92.024110}.
\bibitem{Rossstuffedtitanates}
\bibinfo{author}{K.~A. Ross}, \bibinfo{author}{T.~Proffen},
  \bibinfo{author}{H.~A. Dabkowska}, \bibinfo{author}{J.~A. Quilliam},
  \bibinfo{author}{L.~R. Yaraskavitch}, \bibinfo{author}{J.~B. Kycia}, and
  \bibinfo{author}{B.~D. Gaulin}, \bibinfo{journal}{Phys. Rev. B}
  \bibinfo{volume}{\textbf{86}}, \bibinfo{pages}{174424} (\bibinfo{date}{Nov
  2012}), \url{http://link.aps.org/doi/10.1103/PhysRevB.86.174424}.
\bibitem{Rebuffi}
\bibinfo{author}{L.~Rebuffi}, \bibinfo{author}{J.~R. Plaisier},
  \bibinfo{author}{M.~Abdellatief}, \bibinfo{author}{A.~Lausi}, and
  \bibinfo{author}{P.~Scardi}, \bibinfo{journal}{Zeitschrift für anorganische
  und allgemeine Chemie} \bibinfo{volume}{\textbf{640}}(15),
  \bibinfo{pages}{3100} (\bibinfo{date}{2014}),
  \url{http://dx.doi.org/10.1002/zaac.201400163}.
\bibitem{RODRIGUEZCARVAJAL199355}
\bibinfo{author}{J.~Rodríguez-Carvajal}, \bibinfo{journal}{Physica B:
  Condensed Matter} \bibinfo{volume}{\textbf{192}}(1), \bibinfo{pages}{55 }
  (\bibinfo{date}{1993}),
  \url{http://www.sciencedirect.com/science/article/pii/092145269390108I}.
\bibitem{SUBRAMANIAN198355}
\bibinfo{author}{M.~Subramanian}, \bibinfo{author}{G.~Aravamudan}, and
  \bibinfo{author}{G.~S. Rao}, \bibinfo{journal}{Progress in Solid State
  Chemistry} \bibinfo{volume}{\textbf{15}}(2), \bibinfo{pages}{55 }
  (\bibinfo{date}{1983}),
  \url{http://www.sciencedirect.com/science/article/pii/0079678683900018}.
\bibitem{Chien}
\bibinfo{author}{C.~L. Chien} and \bibinfo{author}{A.~W. Sleight},
  \bibinfo{journal}{Phys. Rev. B} \bibinfo{volume}{\textbf{18}},
  \bibinfo{pages}{2031} (\bibinfo{date}{Sep 1978}),
  \url{https://link.aps.org/doi/10.1103/PhysRevB.18.2031}.
\bibitem{Millican}
\bibinfo{author}{J.~N. Millican}, \bibinfo{author}{R.~T. Macaluso},
  \bibinfo{author}{S.~Nakatsuji}, \bibinfo{author}{Y.~Machida},
  \bibinfo{author}{Y.~Maeno}, and \bibinfo{author}{J.~Y. Chan},
  \bibinfo{journal}{Materials Research Bulletin}
  \bibinfo{volume}{\textbf{42}}(5), \bibinfo{pages}{928 }
  (\bibinfo{date}{2007}),
  \url{http://www.sciencedirect.com/science/article/pii/S0025540806003436}.
\bibitem{KENNEDY1997303}
\bibinfo{author}{B.~J. Kennedy}, \bibinfo{journal}{Physica B: Condensed Matter}
  \bibinfo{volume}{\textbf{241}}, \bibinfo{pages}{303 } (\bibinfo{date}{1997}),
  \url{http://www.sciencedirect.com/science/article/pii/S092145269700570X}.
\bibitem{Lau}
\bibinfo{author}{G.~Lau}, \bibinfo{author}{B.~Muegge},
  \bibinfo{author}{T.~McQueen}, \bibinfo{author}{E.~Duncan}, and
  \bibinfo{author}{R.~Cava}, \bibinfo{journal}{Journal of Solid State
  Chemistry} \bibinfo{volume}{\textbf{179}}(10), \bibinfo{pages}{3126 }
  (\bibinfo{date}{2006}),
  \url{http://www.sciencedirect.com/science/article/pii/S0022459606003392}.
\bibitem{Shannon:a12967}
\bibinfo{author}{R.~D. Shannon}, \bibinfo{journal}{Acta Crystallographica
  Section A} \bibinfo{volume}{\textbf{32}}(5), \bibinfo{pages}{751}
  (\bibinfo{date}{Sep 1976}), \url{https://doi.org/10.1107/S0567739476001551}.
\bibitem{Cordfunke}
\bibinfo{author}{E.~H.~P. Cordfunke} and \bibinfo{author}{G.~Meyer},
  \bibinfo{journal}{Recueil des Travaux Chimiques des Pays-Bas}
  \bibinfo{volume}{\textbf{81}}(6), \bibinfo{pages}{495}
  (\bibinfo{date}{1962}), \url{http://dx.doi.org/10.1002/recl.19620810608}.
\bibitem{Krankendonk}
\bibinfo{author}{J.~VAN~KRANENDONK} and \bibinfo{author}{J.~H. VAN~VLECK},
  \bibinfo{journal}{Rev. Mod. Phys.} \bibinfo{volume}{\textbf{30}},
  \bibinfo{pages}{1} (\bibinfo{date}{Jan 1958}),
  \url{https://link.aps.org/doi/10.1103/RevModPhys.30.1}.
\bibitem{Blacklock}
\bibinfo{author}{K.~Blacklock} and \bibinfo{author}{H.~W. White},
  \bibinfo{journal}{The Journal of Chemical Physics}
  \bibinfo{volume}{\textbf{72}}(3), \bibinfo{pages}{2191}
  (\bibinfo{date}{1980}), \eprint{http://dx.doi.org/10.1063/1.439315},
  \url{http://dx.doi.org/10.1063/1.439315}.
\bibitem{Kittel:ISSP}
\bibinfo{author}{C.~Kittel}, \bibinfo{title}{\emph{{Introduction to Solid State
  Physics}}} (\bibinfo{publisher}{John Wiley \& Sons, Inc.}, New York,
  \bibinfo{year}{1986}), 6th ed.
\bibitem{Donnerer}
\bibinfo{author}{C.~Donnerer}, \bibinfo{author}{M.~C. Rahn},
  \bibinfo{author}{M.~M. Sala}, \bibinfo{author}{J.~G. Vale},
  \bibinfo{author}{D.~Pincini}, \bibinfo{author}{J.~Strempfer},
  \bibinfo{author}{M.~Krisch}, \bibinfo{author}{D.~Prabhakaran},
  \bibinfo{author}{A.~T. Boothroyd}, and \bibinfo{author}{D.~F. McMorrow},
  \bibinfo{journal}{Phys. Rev. Lett.} \bibinfo{volume}{\textbf{117}},
  \bibinfo{pages}{037201} (\bibinfo{date}{Jul 2016}),
  \url{https://link.aps.org/doi/10.1103/PhysRevLett.117.037201}.
\bibitem{Sagayama}
\bibinfo{author}{H.~Sagayama}, \bibinfo{author}{D.~Uematsu},
  \bibinfo{author}{T.~Arima}, \bibinfo{author}{K.~Sugimoto},
  \bibinfo{author}{J.~J. Ishikawa}, \bibinfo{author}{E.~O'Farrell}, and
  \bibinfo{author}{S.~Nakatsuji}, \bibinfo{journal}{Phys. Rev. B}
  \bibinfo{volume}{\textbf{87}}, \bibinfo{pages}{100403} (\bibinfo{date}{Mar
  2013}), \url{https://link.aps.org/doi/10.1103/PhysRevB.87.100403}.
\bibitem{Lef}
\bibinfo{author}{E.~Lefran\ifmmode~\mbox{\c{c}}\else \c{c}\fi{}ois},
  \bibinfo{author}{V.~Simonet}, \bibinfo{author}{R.~Ballou},
  \bibinfo{author}{E.~Lhotel}, \bibinfo{author}{A.~Hadj-Azzem},
  \bibinfo{author}{S.~Kodjikian}, \bibinfo{author}{P.~Lejay},
  \bibinfo{author}{P.~Manuel}, \bibinfo{author}{D.~Khalyavin}, and
  \bibinfo{author}{L.~C. Chapon}, \bibinfo{journal}{Phys. Rev. Lett.}
  \bibinfo{volume}{\textbf{114}}, \bibinfo{pages}{247202} (\bibinfo{date}{Jun
  2015}), \url{https://link.aps.org/doi/10.1103/PhysRevLett.114.247202}.
\bibitem{EuVanVleck}
\bibinfo{author}{K.~Gatterer} and \bibinfo{author}{H.~P. Fritzer},
  \bibinfo{journal}{Journal of Physics: Condensed Matter}
  \bibinfo{volume}{\textbf{4}}(19), \bibinfo{pages}{4667}
  (\bibinfo{date}{1992}), \url{http://stacks.iop.org/0953-8984/4/i=19/a=009}.
\bibitem{Graf}
\bibinfo{author}{M.~J. Graf}, \bibinfo{author}{S.~M. Disseler},
  \bibinfo{author}{C.~Dhital}, \bibinfo{author}{T.~Hogan},
  \bibinfo{author}{M.~Bojko}, \bibinfo{author}{A.~Amato},
  \bibinfo{author}{H.~Luetkens}, \bibinfo{author}{C.~Baines},
  \bibinfo{author}{D.~Margineda}, \bibinfo{author}{S.~R. Giblin},
  \emph{et~al.}, \bibinfo{journal}{Journal of Physics: Conference Series}
  \bibinfo{volume}{\textbf{551}}(1), \bibinfo{pages}{012020}
  (\bibinfo{date}{2014}),
  \url{http://stacks.iop.org/1742-6596/551/i=1/a=012020}.
\bibitem{Macdonald}
\bibinfo{author}{D.~K.~C. MacDonald}, in \emph{Thermoeectricity: An
  interoduction to the principles} (\bibinfo{publisher}{Dover Publication,
  INC.}, \bibinfo{year}{2006}),
  \url{http://store.doverpublications.com/0486453049.html}.
\bibitem{BOUCHARD1971669}
\bibinfo{author}{R.~Bouchard} and \bibinfo{author}{J.~Gillson},
  \bibinfo{journal}{Materials Research Bulletin}
  \bibinfo{volume}{\textbf{6}}(8), \bibinfo{pages}{669 }
  (\bibinfo{date}{1971}),
  \url{http://www.sciencedirect.com/science/article/pii/0025540871901000}.
\bibitem{Minervini}
\bibinfo{author}{L.~Minervini}, \bibinfo{author}{R.~W. Grimes}, and
  \bibinfo{author}{K.~E. Sickafus}, \bibinfo{journal}{Journal of the American
  Ceramic Society} \bibinfo{volume}{\textbf{83}}(8), \bibinfo{pages}{1873}
  (\bibinfo{date}{2000}),
  \url{http://dx.doi.org/10.1111/j.1151-2916.2000.tb01484.x}.
\bibitem{Gomzalez}
\bibinfo{author}{J.~Gonz\'alez}, \bibinfo{journal}{Phys. Rev. B}
  \bibinfo{volume}{\textbf{96}}, \bibinfo{pages}{081104} (\bibinfo{date}{Aug
  2017}), \url{https://link.aps.org/doi/10.1103/PhysRevB.96.081104}.
\bibitem{Ishii}
\bibinfo{author}{F.~Ishii}, \bibinfo{author}{Y.~P. Mizuta},
  \bibinfo{author}{T.~Kato}, \bibinfo{author}{T.~Ozaki},
  \bibinfo{author}{H.~Weng}, and \bibinfo{author}{S.~Onoda},
  \bibinfo{journal}{Journal of the Physical Society of Japan}
  \bibinfo{volume}{\textbf{84}}(7), \bibinfo{pages}{073703}
  (\bibinfo{date}{2015}), \eprint{https://doi.org/10.7566/JPSJ.84.073703},
  \url{https://doi.org/10.7566/JPSJ.84.073703}.
\bibitem{HongbinZhang}
\bibinfo{author}{H.~Zhang}, \bibinfo{author}{K.~Haule}, and
  \bibinfo{author}{D.~Vanderbilt}, \bibinfo{journal}{Phys. Rev. Lett.}
  \bibinfo{volume}{\textbf{118}}, \bibinfo{pages}{026404} (\bibinfo{date}{Jan
  2017}), \url{https://link.aps.org/doi/10.1103/PhysRevLett.118.026404}.
\bibitem{Apetrei}
\bibinfo{author}{A.~Apetrei}, \bibinfo{author}{I.~Mirebeau},
  \bibinfo{author}{I.~Goncharenko}, and \bibinfo{author}{W.~A. Crichton},
  \bibinfo{journal}{Journal of Physics: Condensed Matter}
  \bibinfo{volume}{\textbf{19}}(37), \bibinfo{pages}{376208}
  (\bibinfo{date}{2007}),
  \url{http://stacks.iop.org/0953-8984/19/i=37/a=376208}.
\bibitem{Saha}
\bibinfo{author}{S.~Saha}, \bibinfo{author}{D.~V.~S. Muthu},
  \bibinfo{author}{C.~Pascanut}, \bibinfo{author}{N.~Dragoe},
  \bibinfo{author}{R.~Suryanarayanan}, \bibinfo{author}{G.~Dhalenne},
  \bibinfo{author}{A.~Revcolevschi}, \bibinfo{author}{S.~Karmakar},
  \bibinfo{author}{S.~M. Sharma}, and \bibinfo{author}{A.~K. Sood},
  \bibinfo{journal}{Phys. Rev. B} \bibinfo{volume}{\textbf{74}},
  \bibinfo{pages}{064109} (\bibinfo{date}{Aug 2006}),
  \url{https://link.aps.org/doi/10.1103/PhysRevB.74.064109}.
\bibitem{Krempa2012}
\bibinfo{author}{W.~Witczak-Krempa} and \bibinfo{author}{Y.~B. Kim},
  \bibinfo{journal}{Phys. Rev. B} \bibinfo{volume}{\textbf{85}},
  \bibinfo{pages}{045124} (\bibinfo{date}{Jan 2012}),
  \url{https://link.aps.org/doi/10.1103/PhysRevB.85.045124}.
\bibitem{Sakata}
\bibinfo{author}{M.~Sakata}, \bibinfo{author}{T.~Kagayama},
  \bibinfo{author}{K.~Shimizu}, \bibinfo{author}{K.~Matsuhira},
  \bibinfo{author}{S.~Takagi}, \bibinfo{author}{M.~Wakeshima}, and
  \bibinfo{author}{Y.~Hinatsu}, \bibinfo{journal}{Phys. Rev. B}
  \bibinfo{volume}{\textbf{83}}, \bibinfo{pages}{041102} (\bibinfo{date}{Jan
  2011}), \url{https://link.aps.org/doi/10.1103/PhysRevB.83.041102}.
\bibitem{Young}
\bibinfo{author}{R.~A. Young}, in \emph{The Rietveld Method}
  (\bibinfo{publisher}{Oxford Science Publication}, \bibinfo{year}{1995}),
  \url{https://global.oup.com/academic/product/the-rietveld-method-9780198559122?cc=in&lang=en&#l}.
\bibitem{KENNEDY1996}
\bibinfo{author}{B.~Kennedy} and \bibinfo{author}{T.~Vogt},
  \bibinfo{journal}{Journal of Solid State Chemistry}
  \bibinfo{volume}{\textbf{126}}(2), \bibinfo{pages}{261 }
  (\bibinfo{date}{1996}),
  \url{http://www.sciencedirect.com/science/article/pii/S0022459696903370}.
\bibitem{Sleight}
\bibinfo{author}{C.~L. Chien} and \bibinfo{author}{A.~W. Sleight},
  \bibinfo{journal}{Phys. Rev. B} \bibinfo{volume}{\textbf{18}},
  \bibinfo{pages}{2031} (\bibinfo{date}{Sep 1978}),
  \url{https://link.aps.org/doi/10.1103/PhysRevB.18.2031}.
\bibitem{Bi227thinfilm}
\bibinfo{author}{X.~Y. T. Z. W. K. P. K. C. A. P. L. Y. L. Z. L. H. M. S. M. A.
  J. A. C. G. L. N. J. Q. X. H. J. J. Z. S.~X. Yang, W.~C.},
  \bibinfo{journal}{Scientific Reports} \bibinfo{volume}{\textbf{7740}}
  (\bibinfo{date}{Aug 2017}).
\bibitem{Rebuffi:ZAAC201400163}
\bibinfo{author}{L.~Rebuffi}, \bibinfo{author}{J.~R. Plaisier},
  \bibinfo{author}{M.~Abdellatief}, \bibinfo{author}{A.~Lausi}, and
  \bibinfo{author}{P.~Scardi}, \bibinfo{journal}{Zeitschrift für anorganische
  und allgemeine Chemie} \bibinfo{volume}{\textbf{640}}(15),
  \bibinfo{pages}{3100} (\bibinfo{date}{2014}),
  \url{http://dx.doi.org/10.1002/zaac.201400163}.
\bibitem{trigonal_Uematsu}
\bibinfo{author}{D.~Uematsu}, \bibinfo{author}{H.~Sagayama},
  \bibinfo{author}{T.-h. Arima}, \bibinfo{author}{J.~J. Ishikawa},
  \bibinfo{author}{S.~Nakatsuji}, \bibinfo{author}{H.~Takagi},
  \bibinfo{author}{M.~Yoshida}, \bibinfo{author}{J.~Mizuki}, and
  \bibinfo{author}{K.~Ishii}, \bibinfo{journal}{Phys. Rev. B}
  \bibinfo{volume}{\textbf{92}}, \bibinfo{pages}{094405} (\bibinfo{date}{Sep
  2015}), \url{http://link.aps.org/doi/10.1103/PhysRevB.92.094405}.
\bibitem{topologicalyang}
\bibinfo{author}{B.-J. Yang} and \bibinfo{author}{Y.~B. Kim},
  \bibinfo{journal}{Phys. Rev. B} \bibinfo{volume}{\textbf{82}},
  \bibinfo{pages}{085111} (\bibinfo{date}{Aug 2010}),
  \url{http://link.aps.org/doi/10.1103/PhysRevB.82.085111}.
\bibitem{Mehdikargarian}
\bibinfo{author}{M.~Kargarian}, \bibinfo{author}{J.~Wen}, and
  \bibinfo{author}{G.~A. Fiete}, \bibinfo{journal}{Phys. Rev. B}
  \bibinfo{volume}{\textbf{83}}, \bibinfo{pages}{165112} (\bibinfo{date}{Apr
  2011}), \url{http://link.aps.org/doi/10.1103/PhysRevB.83.165112}.
\bibitem{gamma}
\bibinfo{author}{G.~R. Stewart}, \bibinfo{journal}{Review of Scientific
  Instruments} \bibinfo{volume}{\textbf{54}}(1), \bibinfo{pages}{1}
  (\bibinfo{date}{1983}).
\bibitem{heatcapacity}
\bibinfo{author}{C.~L. Yaws}, \bibinfo{author}{M.~Han}, and
  \bibinfo{author}{S.~D. Sheth}, in \emph{Inorganic Compounds and Elements}
  (\bibinfo{publisher}{Gulf Professional Publishing}, \bibinfo{year}{1996}),
  \bibinfo{volume}{vol.~4 of \emph{Handbook of Thermodynamic Diagrams}},
  \bibinfo{pages}{pp. 357 -- 362},
  \url{http://www.sciencedirect.com/science/article/pii/B9780884158608500374}.
\bibitem{EuVan2}
\bibinfo{author}{M.~Tovar}, \bibinfo{author}{D.~Rao},
  \bibinfo{author}{J.~Barnett}, \bibinfo{author}{S.~B. Oseroff},
  \bibinfo{author}{J.~D. Thompson}, \bibinfo{author}{S.-W. Cheong},
  \bibinfo{author}{Z.~Fisk}, \bibinfo{author}{D.~C. Vier}, and
  \bibinfo{author}{S.~Schultz}, \bibinfo{journal}{Phys. Rev. B}
  \bibinfo{volume}{\textbf{39}}, \bibinfo{pages}{2661} (\bibinfo{date}{Feb
  1989}), \url{http://link.aps.org/doi/10.1103/PhysRevB.39.2661}.
\bibitem{KOO1998269}
\bibinfo{author}{H.-J. Koo}, \bibinfo{author}{M.-H. Whangbo}, and
  \bibinfo{author}{B.~Kennedy}, \bibinfo{journal}{Journal of Solid State
  Chemistry} \bibinfo{volume}{\textbf{136}}(2), \bibinfo{pages}{269 }
  (\bibinfo{date}{1998}),
  \url{http://www.sciencedirect.com/science/article/pii/S0022459697977057}.
\bibitem{DisselerYb}
\bibinfo{author}{S.~M. Disseler}, \bibinfo{author}{C.~Dhital},
  \bibinfo{author}{A.~Amato}, \bibinfo{author}{S.~R. Giblin},
  \bibinfo{author}{C.~de~la Cruz}, \bibinfo{author}{S.~D. Wilson}, and
  \bibinfo{author}{M.~J. Graf}, \bibinfo{journal}{Phys. Rev. B}
  \bibinfo{volume}{\textbf{86}}, \bibinfo{pages}{014428} (\bibinfo{date}{Jul
  2012}), \url{http://link.aps.org/doi/10.1103/PhysRevB.86.014428}.
\bibitem{Taira}
\bibinfo{author}{N.~Taira}, \bibinfo{author}{M.~Wakeshima}, and
  \bibinfo{author}{Y.~Hinatsu}, \bibinfo{journal}{Journal of Physics: Condensed
  Matter} \bibinfo{volume}{\textbf{13}}(23), \bibinfo{pages}{5527}
  (\bibinfo{date}{2001}), \url{http://stacks.iop.org/0953-8984/13/i=23/a=312}.
\bibitem{Hastings:a24110}
\bibinfo{author}{J.~B. Hastings}, \bibinfo{author}{W.~Thomlinson}, and
  \bibinfo{author}{D.~E. Cox}, \bibinfo{journal}{Journal of Applied
  Crystallography} \bibinfo{volume}{\textbf{17}}(2), \bibinfo{pages}{85}
  (\bibinfo{date}{Apr 1984}), \url{https://doi.org/10.1107/S0021889884011043}.
\bibitem{Yanagishima}
\bibinfo{author}{D.~Yanagishima} and \bibinfo{author}{Y.~Maeno},
  \bibinfo{journal}{Journal of the Physical Society of Japan}
  \bibinfo{volume}{\textbf{70}}(10), \bibinfo{pages}{2880}
  (\bibinfo{date}{2001}), \url{http://dx.doi.org/10.1143/JPSJ.70.2880}.
\bibitem{Taniguchi}
\bibinfo{author}{T.~Taniguchi}, \bibinfo{author}{H.~Kadowaki},
  \bibinfo{author}{H.~Takatsu}, \bibinfo{author}{B.~F\aa{}k},
  \bibinfo{author}{J.~Ollivier}, \bibinfo{author}{T.~Yamazaki},
  \bibinfo{author}{T.~J. Sato}, \bibinfo{author}{H.~Yoshizawa},
  \bibinfo{author}{Y.~Shimura}, \bibinfo{author}{T.~Sakakibara}, \emph{et~al.},
  \bibinfo{journal}{Phys. Rev. B} \bibinfo{volume}{\textbf{87}},
  \bibinfo{pages}{060408} (\bibinfo{date}{Feb 2013}),
  \url{http://link.aps.org/doi/10.1103/PhysRevB.87.060408}.
\bibitem{Disseler2014}
\bibinfo{author}{S.~M. Disseler}, \bibinfo{journal}{Phys. Rev. B}
  \bibinfo{volume}{\textbf{89}}, \bibinfo{pages}{140413} (\bibinfo{date}{Apr
  2014}), \url{http://link.aps.org/doi/10.1103/PhysRevB.89.140413}.
\bibitem{Zhu1hole1doping}
\bibinfo{author}{W.~K. Zhu}, \bibinfo{author}{M.~Wang},
  \bibinfo{author}{B.~Seradjeh}, \bibinfo{author}{F.~Yang}, and
  \bibinfo{author}{S.~X. Zhang}, \bibinfo{journal}{Phys. Rev. B}
  \bibinfo{volume}{\textbf{90}}, \bibinfo{pages}{054419} (\bibinfo{date}{Aug
  2014}), \url{http://link.aps.org/doi/10.1103/PhysRevB.90.054419}.
\bibitem{Ueda1pressureMIT}
\bibinfo{author}{K.~Ueda}, \bibinfo{author}{J.~Fujioka},
  \bibinfo{author}{C.~Terakura}, and \bibinfo{author}{Y.~Tokura},
  \bibinfo{journal}{Phys. Rev. B} \bibinfo{volume}{\textbf{92}},
  \bibinfo{pages}{121110} (\bibinfo{date}{Sep 2015}),
  \url{http://link.aps.org/doi/10.1103/PhysRevB.92.121110}.
\bibitem{Prandopressure}
\bibinfo{author}{G.~Prando}, \bibinfo{author}{R.~Dally},
  \bibinfo{author}{W.~Schottenhamel}, \bibinfo{author}{Z.~Guguchia},
  \bibinfo{author}{S.-H. Baek}, \bibinfo{author}{R.~Aeschlimann},
  \bibinfo{author}{A.~U.~B. Wolter}, \bibinfo{author}{S.~D. Wilson},
  \bibinfo{author}{B.~B\"uchner}, and \bibinfo{author}{M.~J. Graf},
  \bibinfo{journal}{Phys. Rev. B} \bibinfo{volume}{\textbf{93}},
  \bibinfo{pages}{104422} (\bibinfo{date}{Mar 2016}),
  \url{http://link.aps.org/doi/10.1103/PhysRevB.93.104422}.
\bibitem{Kanno1993106}
\bibinfo{author}{R.~Kanno}, \bibinfo{author}{Y.~Takeda},
  \bibinfo{author}{T.~Yamamoto}, \bibinfo{author}{Y.~Kawamoto}, and
  \bibinfo{author}{O.~Yamamoto}, \bibinfo{journal}{Journal of Solid State
  Chemistry} \bibinfo{volume}{\textbf{102}}(1), \bibinfo{pages}{106 }
  (\bibinfo{date}{1993}),
  \url{//www.sciencedirect.com/science/article/pii/S0022459683710121}.
\bibitem{Pesin2010376}
\bibinfo{author}{D.~Pesin} and \bibinfo{author}{L.~Balents},
  \bibinfo{journal}{Nature Physics} \bibinfo{volume}{\textbf{6}},
  \bibinfo{pages}{376 } (\bibinfo{date}{2010}),
  \url{http://www.nature.com/nphys/journal/v6/n5/abs/nphys1606.html}.
\bibitem{Sushkov}
\bibinfo{author}{A.~B. Sushkov}, \bibinfo{author}{J.~B. Hofmann},
  \bibinfo{author}{G.~S. Jenkins}, \bibinfo{author}{J.~Ishikawa},
  \bibinfo{author}{S.~Nakatsuji}, \bibinfo{author}{S.~Das~Sarma}, and
  \bibinfo{author}{H.~D. Drew}, \bibinfo{journal}{Phys. Rev. B}
  \bibinfo{volume}{\textbf{92}}, \bibinfo{pages}{241108} (\bibinfo{date}{Dec
  2015}), \url{http://link.aps.org/doi/10.1103/PhysRevB.92.241108}.
\bibitem{NakatsujiPr227SpinLiq}
\bibinfo{author}{S.~Nakatsuji}, \bibinfo{author}{Y.~Machida},
  \bibinfo{author}{Y.~Maeno}, \bibinfo{author}{T.~Tayama},
  \bibinfo{author}{T.~Sakakibara}, \bibinfo{author}{J.~v. Duijn},
  \bibinfo{author}{L.~Balicas}, \bibinfo{author}{J.~N. Millican},
  \bibinfo{author}{R.~T. Macaluso}, and \bibinfo{author}{J.~Y. Chan},
  \bibinfo{journal}{Phys. Rev. Lett.} \bibinfo{volume}{\textbf{96}},
  \bibinfo{pages}{087204} (\bibinfo{date}{Mar 2006}),
  \url{http://link.aps.org/doi/10.1103/PhysRevLett.96.087204}.
\bibitem{MacLaughlinPr227stuffing}
\bibinfo{author}{D.~E. MacLaughlin}, \bibinfo{author}{O.~O. Bernal},
  \bibinfo{author}{L.~Shu}, \bibinfo{author}{J.~Ishikawa},
  \bibinfo{author}{Y.~Matsumoto}, \bibinfo{author}{J.-J. Wen},
  \bibinfo{author}{M.~Mourigal}, \bibinfo{author}{C.~Stock},
  \bibinfo{author}{G.~Ehlers}, \bibinfo{author}{C.~L. Broholm}, \emph{et~al.},
  \bibinfo{journal}{Phys. Rev. B} \bibinfo{volume}{\textbf{92}},
  \bibinfo{pages}{054432} (\bibinfo{date}{Aug 2015}),
  \url{http://link.aps.org/doi/10.1103/PhysRevB.92.054432}.
\bibitem{Zhang}
\bibinfo{author}{F.~X. Zhang}, \bibinfo{author}{M.~Lang},
  \bibinfo{author}{Z.~Liu}, and \bibinfo{author}{R.~C. Ewing},
  \bibinfo{journal}{Phys. Rev. Lett.} \bibinfo{volume}{\textbf{105}},
  \bibinfo{pages}{015503} (\bibinfo{date}{Jun 2010}),
  \url{https://link.aps.org/doi/10.1103/PhysRevLett.105.015503}.
\bibitem{Prando}
\bibinfo{author}{G.~Prando}, \bibinfo{author}{R.~Dally},
  \bibinfo{author}{W.~Schottenhamel}, \bibinfo{author}{Z.~Guguchia},
  \bibinfo{author}{S.-H. Baek}, \bibinfo{author}{R.~Aeschlimann},
  \bibinfo{author}{A.~U.~B. Wolter}, \bibinfo{author}{S.~D. Wilson},
  \bibinfo{author}{B.~B\"uchner}, and \bibinfo{author}{M.~J. Graf},
  \bibinfo{journal}{Phys. Rev. B} \bibinfo{volume}{\textbf{93}},
  \bibinfo{pages}{104422} (\bibinfo{date}{Mar 2016}),
  \url{https://link.aps.org/doi/10.1103/PhysRevB.93.104422}.

\end{thebibliography}
\bibliographystyle{revtex}

\pagebreak
\end{document}